\documentclass[journal]{IEEEtran}

\usepackage{amsmath,amsfonts,amsthm,amssymb,epsfig,amssymb,bm}
\usepackage{titlesec}
\usepackage{tcolorbox}
\usepackage{textcomp}
\usepackage{algorithmic}
\usepackage{algorithm}
\usepackage{array}
\usepackage{multirow}
\usepackage{indentfirst}
\usepackage{stfloats}
\usepackage{url}
\usepackage{verbatim}
\usepackage{graphicx}
\usepackage{cite}
\usepackage{caption}
\usepackage{float} 
\usepackage{pifont}
\newtheorem{definition}{Definition}
\newtheorem{theorem}{Theorem}
\newtheorem{remark}{Remark}
\usepackage{balance}

\usepackage[caption=false,font=normalsize,labelfont=sf,textfont=sf]{subfig}
\usepackage{bbding}

\usepackage{hyperref} % 用于链接
\begin{document}

\title{Coalition Formation for Heterogeneous Federated Learning Enabled Channel Estimation in RIS-assisted Cell-free
MIMO}

	\author{
		Nan Qi, ~\IEEEmembership{Senior Member,~IEEE,} Haoxuan Liu, ~\IEEEmembership{Member,~IEEE,} Theodoros A. Tsiftsis, ~\IEEEmembership{Senior Member,~IEEE,} Alexandros-Apostolos A. Boulogeorgos, ~\IEEEmembership{Senior Member,~IEEE,} Fuhui Zhou, ~\IEEEmembership{Senior Member,~IEEE,}  Shi Jin, ~\IEEEmembership{Fellow,~IEEE}, Qihui Wu, ~\IEEEmembership{Fellow,~IEEE}
		%%	Michael~Shell,~\IEEEmembership{Member,~IEEE,}
		%%        John~Doe,~\IEEEmembership{Fellow,~OSA,}
		%%        and~Jane~Doe,~\IEEEmembership{Life~Fellow,~IEEE}% <-this % stops a space
	\IEEEcompsocitemizethanks{\IEEEcompsocthanksitem This article was accepted in part at the IEEE Wireless Communications and Networking Conference(WCNC), Milan, Italy, March 24–27, 2025\cite{WCNC}. 
    \IEEEcompsocthanksitem Nan Qi, Haoxuan Liu, Fuhui Zhou, and Qihui Wu are with the Key Laboratory of Dynamic Cognitive System of Electromagnetic Spectrum Space, Ministry of Industry and Information Technology, Nanjing University of Aeronautics and Astronautics, Nanjing, China, 210016. E-mail: nanqi.commun@gmail.com, nuaalhx@nuaa.edu.cn, zhoufuhui@ieee.org, wuqihui@nuaa.edu.cn. 
        \IEEEcompsocthanksitem Theodoros A. Tsiftsis is with the Department of Informatics and Telecommunications, University of Thessaly, 35100 Lamia, Greece, E-mail: tsiftsis@uth.gr.
        \IEEEcompsocthanksitem Alexandros -Apostolos. A. Boulogeorgos is with the Department of Electrical and Computer Engineering, University of Western Macedonia, 50100 Kozani, Greece, E-mail: aboulogeorgos@uowm.gr.        
        \IEEEcompsocthanksitem S. Jin is with the National Mobile Communications Research Laboratory and Frontiers Science Center for Mobile Information Communication and Security, Southeast University, Nanjing 210096, P.R. China, E-mail: jinshi@seu.edu.cn.
			 			
			}% <-this % stops an unwanted space
						% \thanks{This work was supported by the National Key R\&D Program of China (No. Grant 2018YFB1800801), the National Natural Science Foundation of China under Key Project (No. 61931011), the National Natural Science Foundation of China (No. 61827801, 61801218, 61901523, 62071223, and 62031012), the Fundamental Research Funds for the Central Universities, NO. NT2021017, and Young Elite Scientist Sponsorship Program by CAST. The work was supported in part by ERA-NET Smart Energy Systems SG+ 2017 Program, ”SMART-MLA” with Project number 89029 (and SWEA number 42811-2), in part by FORMAS project entitled ”Intelligent Energy Management in Smart Community with Distributed Machine Learning”, number 2021-00306, and in part by Swedish Research Council Project entitled ”Coding for Large-scale Distributed Machine Learning”, number 2021-04772”. (Corresponding author: Nan Qi)}
                        }

\markboth{Journal of \LaTeX\ Class Files,~Vol.~14, No.~8, August~2021}%
	{Shell \MakeLowercase{\textit{et al.}}: Bare Demo of IEEEtran.cls for Computer Society Journals}

\IEEEtitleabstractindextext{%
		\begin{abstract}
            Downlink channel estimation remains a significant bottleneck in reconfigurable intelligent surface-assisted cell-free multiple-input multiple-output communication systems. Conventional approaches primarily  rely on centralized deep learning methods to estimate the high-dimensional and complex cascaded channels. These methods require data aggregation from all users for centralized model training, leading to excessive communication overhead and significant data privacy concerns. Additionally, the large size of local learning models imposes heavy computational demands on end users, necessitating strong computational capabilities that most commercial devices lack. 
            To address the aforementioned challenges, a coalition-formation-guided heterogeneous federated learning (FL) framework is proposed. This framework leverages coalition formation to guide the formation of heterogeneous FL user groups for efficient channel estimation.    Specifically, by utilizing a distributed deep reinforcement learning (DRL) approach,  each FL user   intelligently and independently decides whether to join or leave a coalition, aiming at improving channel estimation accuracy, while reducing local model size and computational costs for end  users. 
            Moreover, to accelerate the    DRL-FL convergence process and  reduce computational burdens on end users, a transfer learning method is introduced. 
            This method incorporates both received reference signal power and distance similarity metrics, by 
            considering that nodes with similar distances to the base station and comparable  received signal power have a strong likelihood of experiencing similar channel fading. Massive experiments performed that reveal that, compared with the benchmarks, the proposed framework significantly reduces the computational overhead of end users by 16\%, improves data privacy, and improves channel estimation accuracy by 20\%.     
		\end{abstract}
		
		}

	\maketitle
	
	\IEEEdisplaynontitleabstractindextext
	\IEEEpeerreviewmaketitle

\section{Introduction}
Over the past decades, wireless technologies have been substantially evolving to meet the challenges of ultra-high capacity, ultra-low latency, and massive device access brought about by fifth-generation (5G) networks and beyond. One of the most powerful technologies is  multiple-input multiple-output (MIMO) \cite{1-MIMO,3-MIMO,4-MIMO}. MIMO networks, comprised of multiple multi-antenna base stations (BSs) serving multiple users within a designated area, now play a pivotal role in enhancing spectral efficiency and network capacity.
However, as the number of mobile communication users surges suddenly, the communication rate of individual users (especially the edge users) will be forced to decrease. 
To enhance system spectral efficiency and quality of service of  edge users, a cell-free MIMO network architecture has recently been proposed\cite{SEMIMO}. In this scheme, all BSs are connected to a central processing unit (CPU), which operates them as a unified MIMO network without cell boundaries. This enables coherent transmission and reception, allowing the BSs to serve all users over the same time-frequency resources by applying spatial multiplexing techniques \cite{name15}.

% .
% Massive MIMO, however, requires the use of expensive and energy-intensive gear, such as high-resolution digital-to-analog converters (DACs) and specialized power amplifiers. 
% On the other hand, a cell-free MIMO network architecture enables all access points (APs) to work together as a MIMO network without cell borders by integrating them all through a central processing unit (CPU).
% By enabling coherent transmission and reception through spatial multiplexing, this configuration optimizes spectrum utilization by offering all users simultaneous services utilizing the same time-frequency resources \cite{name14,name15}.

Meanwhile, reconfigurable intelligent surface/intelligent reflection surfaces (RISs / IRSs) have attracted extensive attention in both academia and industry. RISs can manipulate electromagnetic parameters like phase, amplitude, frequency, and polarization\cite{name1} to create a favorable wireless communication environment, which is considered to be a key technology in the sixth generation (6G)\cite{5-Soc}.
Researchers  have investigated the fundamental constraints and system architecture, revealing that RIS-assisted cell-free  MIMO can greatly increase user communication rates\cite{9-MIMO,name6}. 

Several studies highlight the significance of RIS in wireless networks optimize wireless network performance by simultaneously optimizing the passive beamforming and BS transmit beamforming\cite{10-kw,11-fair,12-kw,13-liu,14-kw}. In particular, the authors in\cite{10-kw} optimized beamforming by weighted minimize mean-square error (WMMSE) maximizing a wireless network user's weighted sum rate (WSR), and the complexity of the algorithm is greatly reduced by deriving closed-form solutions. Furthermore, to optimize the minimum possible rate for cell-free MIMO networks supported by numerous RISs, \cite{11-fair} and \cite{12-kw} have focused on the fairness problem. The authors of \cite{13-liu} and \cite{14-kw} have optimized the energy efficiency of RIS-assisted networks through network deployment as well as beamforming. Nevertheless, prior research is primarily available on the assumption that perfect channel information exists. In massive MIMO, where the electromagnetic environment is intricate and cluttered, obtaining perfect channel state information remains challenging. 

\begin{table*}
% \tiny
\caption{A comparison with existing literature}
\label{tab_1}
\resizebox{1.0\linewidth}{!}{
\begin{tabular}{|c|c|c|c|c|c|c|c|c|}
\hline
& \cite{MMSE,MF,CS,DL_1,DNN-1} & \cite{DML1,DML} & \cite{FL_Elbir, FL_Shen_Qin,FL_Bin} & \cite{HFL1,HFL2}  & \cite{Coalition_1,Coalition_2} & \cite{CF_CE_1} & \cite{RL_CL_1,RL_CL_2,RL_CL_3} & \textbf{Our work} \\
\hline
% & [17-21] & [22-23] & [24-27] & [28-29]  & [30-31] & [32] & [33-35] & \textbf{Our work} \\
% \hline
Cascaded channel  & \checkmark & \checkmark & \checkmark  & \ding{53} & \ding{53} & \checkmark & \ding{53} &\checkmark \\
\hline
Distributed training  & \ding{53} & \checkmark  & \checkmark & \checkmark & \checkmark & \ding{53} & \ding{53} &\checkmark \\
\hline
Coalition formation game  & \ding{53} & \ding{53} & \ding{53} & \ding{53} & \checkmark  & \checkmark & \checkmark & \checkmark \\
\hline
Distributed intelligent decision-making & \ding{53} & \ding{53} & \ding{53} & \ding{53} & \ding{53} & \checkmark & \checkmark &\checkmark \\
\hline
Heterogeneous training model& \ding{53} & \ding{53} & \ding{53} & \checkmark & \ding{53} & \ding{53} & \ding{53} & \checkmark \\
\hline
\end{tabular}
}\vspace{-1.5em} \label{Table_comp}
\end{table*}

\vspace{-1.2em}\subsection{Related Works}\label{relatedworks}
Existing studies, as summarized in TABLE \ref{Table_comp}, have explored channel estimation in RIS-assisted wireless systems, including least square (LS), minimum mean square error (MMSE) \cite{MMSE}, matrix factorization \cite{MF}, and compressed sensing \cite{CS} methods. Despite the great potential of these techniques, the high demands on the antenna array response (e.g., sparsity) and the high complexity of the required computations greatly restrict their practical applications. To address these issues, data-driven methodologies, notably deep learning (DL) algorithms \cite{DL_1}, have emerged as viable alternatives for dealing with the complex non-linear correlations inherent in receiving signal samples. In particular, \cite{DNN-1} employed pilot design and non-local attention modules in neural networks by gradually pruning less significant neurons, reducing the pilot transmission overhead, and improving channel estimation accuracy.

The performance of DL  generally scales with both the model size and the volume of training data. However, large models and high-volume datasets impose heavy computational demands on a single node, often exceeding  its  computation capability.  Distributed machine learning (DML) is a powerful solution to address this challenge  by leveraging parallel processing\cite{DML1}. It divides data into batches or partitions the model across multiple compute nodes for concurrent training of DL models, followed by inter-node communication to update model parameters. 
The research in \cite{DML} applied DML to downlink channel estimation in RIS-based networks and proposed a hierarchical neural network-based DML approach. This method improves estimation accuracy by pre-classifying user groups under different scenarios and then training scenario-specific models using the corresponding data. However,  in DML method, the transmission processes involved in data distribution and model updates  lead to increased   communication overhead and   a risk of data leakage, thereby threatening user privacy. % Furthermore, training accuracy on non-independent and identically distributed (non-IID) datasets remains to be improved. 

\emph{To address the challenges of high communication overhead and privacy leakage, federated learning (FL) schemes, as special kinds of DML, have been recently proposed. In FL, rather than transmitting the entire dataset, only model updates, such as the gradients of model parameters, are exchanged.} There have been some important works on how to utilize FL for RIS channel estimation \cite{FL_Elbir, FL_Shen_Qin}. Elbir \textit{et. al.} applied FL to channel estimation for RIS-assisted MIMO systems\cite{FL_Elbir}. In comparison with  the centralized learning method, the ChannelNet learning method in \cite{FL_Elbir} significantly lowers the  amount of communication overhead by utilizing the FL method. However, the performance, while approaching that of MMSE, remains suboptimal due to the aggregation of parameter updates from all participating users. Compared to \cite{FL_Elbir}, \cite{FL_Shen_Qin} achieves high-precision channel estimation while reducing the number of model parameters by integrating a deep residual network into the FL framework, thereby further decreasing the communication overhead. 
In addition, \cite{FL_Bin} employs hierarchical residual networks and federated learning to classify user distribution scenarios, addressing the decline in channel estimation accuracy caused by uneven user location distribution.
Moreover, the large size of the local learning models places a significant computational burden on end-user devices, requiring powerful computing capabilities that exceed the resources available on many existing devices. \emph{To adapt to the different computing and storage capabilities of the end-user, the network model needs to be flexibly resizable, and heterogeneous federated learning (HFL) is ideally suited to address the requirement.} The authors of \cite{HFL1} articulated customized model aggregation strategies to cope with different types of heterogeneity and  increase the feasibility. Data-free knowledge distillation techniques were introduced to HFL, enabling the efficient transfer of model knowledge without requiring access to raw data. This approach is particularly beneficial in resource-constrained environments  \cite{HFL2}.
%这里写出异构联邦的需求%% 为什么用联盟解决FL组合，和相关研究

In multi-user RIS-MIMO networks, channel attenuation varies significantly among geographically dispersed users. This variability necessitates partitioning users into multiple federated learning (FL) groups, with each group aggregating its own channel model. 
However, \emph{most existing FL applications in channel estimation do not address the optimization of user grouping; instead, they use predefined FL groups. This approach  may deteriorate channel estimation accuracy in the aggregated models.}
A coalition formation game provides a framework for multiple agents (players) to cooperate and share resources (e.g., local model parameters of neural networks) to form coalitions for mutual benefits or shared goals (e.g., improved channel estimation accuracy). This framework supports user group partitioning,   allowing players to form coalitions with others to improve their outcomes  and determining  how the overall benefits, such as rewards  or utility, are distributed among members. \emph{Therefore, the problem of federated learning user grouping can be readily converted into a coalition formation game.} However, few studies have addressed the FL user grouping problem by using coalition formation games.
In \cite{Coalition_1} and \cite{Coalition_2}, coalition formation was employed to quantify user data contribution \cite{Coalition_1}, and prevent user selfishness \cite{Coalition_2}. However, all the above works focus on a single FL aggregated model. Specifically,  the work in \cite{Coalition_1} was the first to incorporate coalition games into personalized federated learning, utilizing Shapley values to assess the contribution of other users' models to an individual model, subsequently selecting parameters that enhance performance and mitigate degradation from malicious users. Literature The authors of \cite{Coalition_2} introduced a coalition member trust mechanism to prevent selfish devices from harming their coalition, mitigating the problems of non-independent and identically distributed (I.I.D) data and communication bottlenecks. 
Moreover, the authors of \cite{CF_CE_1} investigated the trade-off between channel estimation accuracy and communication overhead in centralized computation using clustering algorithms and proved the algorithm's stable convergence. This   reveals the feasibility of combining coalition formation games with federated learning for channel estimation.

%如何寻找联盟稳定结构 问题没闭式的目标表达式
To address the FL user grouping problem through coalition formation games, two subproblems need to be tackled: 1) Establishing   stable coalitions  is essential: This involves  designing an overall payoff function, a payoff distribution policy, and a coalition formation order that align to improve channel estimation accuracy, while ensuring the existence of a stable coalition partition. 2) The identification of a stable coalition partition. 
In particular, for subproblem 2),  the coalition formation of the FL user 
is mathematically intractable  under an unknown channel model, especially when coupled with the iterative framework of federated learning. This makes it infeasible to formulate the overall coalition payoff  and utility   distribution among coalition members, both of which are essential for guiding coalition formation. 
Conventional   approaches, such as mixed-integer programming and heuristic algorithms (e.g., ant colony optimization, genetic algorithms), are no longer applicable.  In order to tackle those problems, a data-driven deep reinforcement learning (DRL) approach can be considered to address the coalition formation problem. 
The authors of \cite{RL_CL_1} introduced the DRL-based coalition formation solution for peer-to-peer energy trading problem, aiming to minimize the cost of energy trading. In addition, coalition formation games are also applied in multi-agent task allocation to model task preferences of different nodes, enhancing task completion efficiency through collaboration among multiple coalitions \cite{RL_CL_2,RL_CL_3}.

\vspace{-1.2em}\subsection{Motivations and Contributions}
Motivated by the above observations and analyses, this paper proposes   
a coalition-formation-guided heterogeneous FL approach for channel estimation in RIS-assisted cell-free MIMO systems. Specifically, the FL user grouping problem is modeled as a coalition formation problem, and the existence of a stable coalition formation solution is theoretically analyzed. \textit{To the best of our knowledge,  this is the first
work that proposes a coalition-formation-guided FL method for channel estimation in RIS-MIMO, which can be readily applied to MIMO networks.}
Moreover,  achieving stable coalition formation solutions is particularly challenging. To address this challenge,  we will propose a distributed DRL method that enables each end user to dependently decide whether to leave or join a coalition namely, a FL user group,  thereby alleviating the burden of a central decision node.  
% We proposed a coalition formation-based user grouping method for FL and developed a distributed Deep Q-Network (DQN), i.e., a Qmix-based algorithm is designed to seek multiple FL user coalition solutions.  Under the FL framework, by limiting information exchange to model parameters, the proposed framework effectively reduces data transmission overhead, enhances data privacy, and ensures high channel estimation accuracy.
To reduce the size of local models for individual users that have very limited computational resources, such as smartwatches or tablets, a heterogeneous FL (HFL)-based approach will be proposed.
Lastly,  to accelerate the convergence of the DRL-federated learning process, further reducing computational burdens on end users, we design a transfer learning method. 

The major contributions of this work are summarized as follows
\begin{itemize}
    \item  This is the first work to 
 utilize coalition formation to enhance FL user grouping and significantly improve channel estimation accuracy.  A coalition altruistic order is proposed, and it is proven that the formulated game has at least one stable coalition partition solution that (at least locally) minimizes the overall channel estimation mean square error (MSE). 
    \item DRL framework, namely,  Qmix-based reinforcement learning, is proposed to seek for the stable coalition formation solution. The distributed framework facilitates autonomy and collaboration among the users in the coalition, where each user can independently make decisions on joining or leaving one coalition based on local information, while taking into account the interests of the whole coalition.  
    It significantly reduces the information exchange overhead, compared with the centralized DQN approach, but only brings slight channel estimation accuracy loss (for example, approximately 2.5\% loss at a 5dB signal-to-noise ratio).
    \item A HFL-based approach is proposed to reduce the size of local neural network models for individual users with very limited computational resources, leading to a substantial decrease in computational and power costs during the training process compared to homogeneous FL. While heterogeneous FL may result in a slight reduction in accuracy, it significantly lowers computational costs (measured in Flops) by 16\% and complexity.  More importantly, under the HFL framework, only a part of model parameters are transmitted, reducing the quantity of data that models transmit, thus improving data privacy compared with that of homogeneous FL. % We identify the   heterogeneous FL (no coalition formation is utilized) and DQN-enabled homogeneous FL as baselines representing the upper and lower performance bounds, respectively. It outperforms the baseline of  lower performance bound by 25\% , while losing less than 4\% accuracy compared with the baseline of  lower performance bound.

    \item  To accelerate the convergence of HFL, we propose a transfer learning approach, referred to as ``FL+RSRP+Dis", which incorporates both received signal reference power (RSRP) and distance similarity metrics to enhance the initialization of local model parameters.
    % Numerical results demonstrate that, compared to ``DML", conventional ``FL", and the ``MMSE algorithm", the ``FL+RSRP+Dis" method improves channel estimation accuracy by at least 30\%.
\end{itemize}
\vspace{-1.5em}\subsection{Organizations}
The RIS-assisted wireless communication system is first introduced in the paper, followed by the problem formulation. Then, a   deep learning method for downlink channel estimation is subsequently presented. Section II discusses the system model along with the foundational models. In Section III, a coalition game-enabled federated learning method is proposed to address the channel estimation problem. Section IV presents numerical results and analyzes convergence and complexity. Concluding remarks are given in Section V.
\begin{figure}[h]
    \centering
    \includegraphics[width=0.70\linewidth]{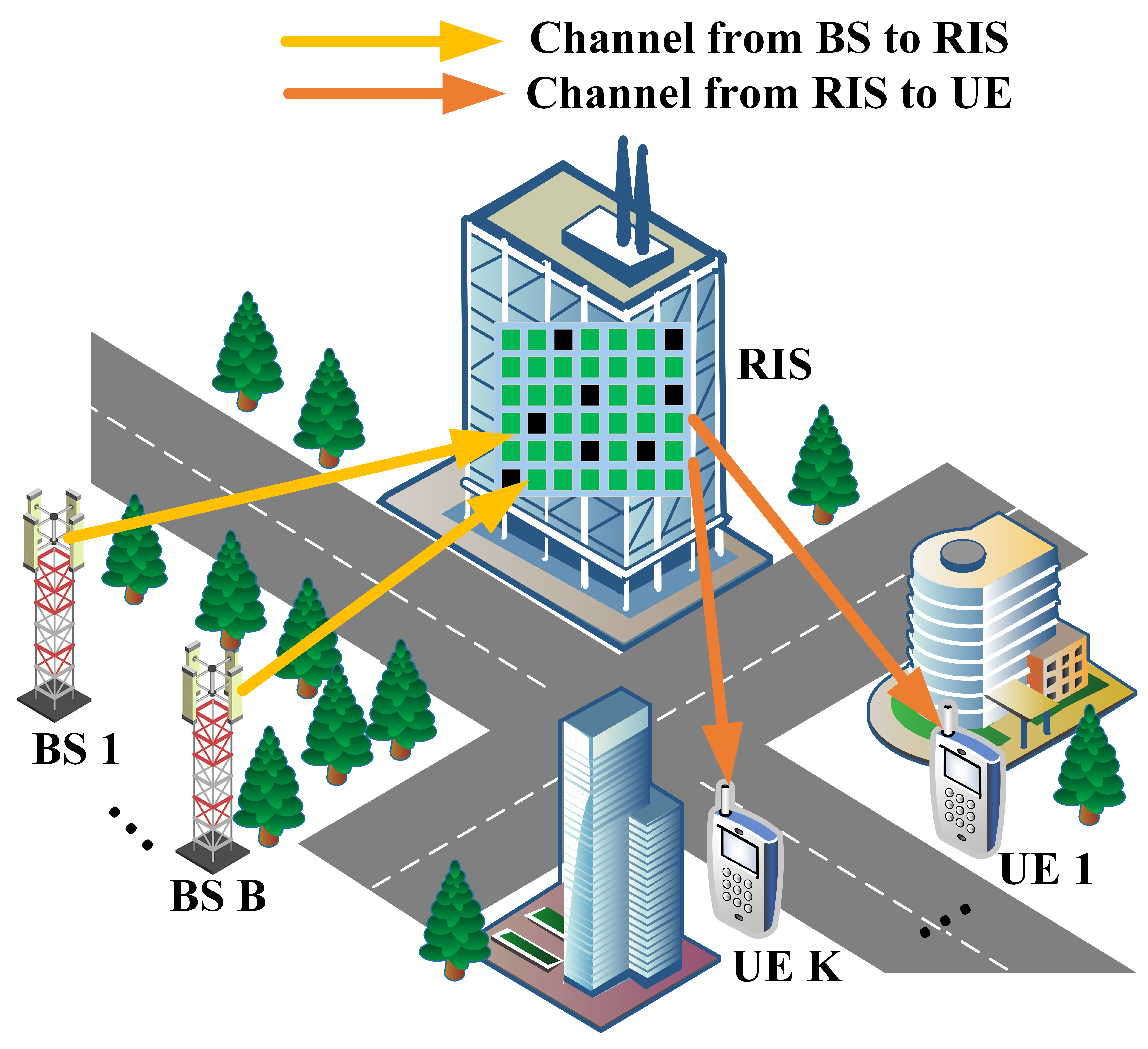}
    \caption{RIS-assisted cell-free MIMO network.}
    \label{System_Model_for_CS}\vspace{-1.5em}
\end{figure}
\section{System Model and Problem Formulation}
% Figure+Description
An RIS-assisted cell-free MIMO wireless downlink communication system is considered, as illustrated in Figure \ref{System_Model_for_CS}. This system comprises $B$ BSs and $K$ user equipment (UEs). Each BS is equipped with $N_t$ antennas, while each  UE is equipped with a single antenna. A single RIS  comprising $N$ elements is deployed between BSs and UEs to enhance the signal quality. All BSs jointly serve each UE, with the beamforming vector from the BSs to UE $k$   denoted by $\mathbf{F}_{k} = [\mathbf{f}_{1,k},...\mathbf{f}_{b,k},...,\mathbf{f}_{B,k}]^{T} \in\mathbb{C}^{\left(N_t \times B\right) \times 1}$, where $\mathbf{f}_{b,k}\in\mathbb{C}^{N_t \times 1}$ represents the beamforming vector from BS $b$ to UE $k$. The phase shift matrix of RIS can be expressed as $\boldsymbol{\Phi}=\operatorname{diag}\left(\mathrm{e}^{j \theta_{1}}, \mathrm{e}^{j \theta_{2}}, \ldots, \mathrm{e}^{j \theta_{N}}\right) \in \mathbb{C}^{N \times N}$, where $\theta_{n}$ is the phase-shift coefficient of the $n$-th RIS element.
\vspace{-1em}\subsection{Channel Modeling}
The channel matrix between BS $b$ to RIS and the channel matrix between RIS to UE $k$ are denoted by $\mathbf{G}_{b} \in \mathbb{C}^{N \times N_{t}}, \mathbf{v}_{k}^{H} \in \mathbb{C}^{1 \times N}$, respectively. The channel matrices can be represented by the widely used Saleh-Valenzuela channel model. Specifically, $\mathbf{G}_{b}$ is given by 
\begin{equation}\vspace{-0.5em}
\begin{small}
\begin{aligned}
\mathbf{G}_b=\sqrt{\frac{N^t N}{L_G}} \sum_{l_1=1}^{L_G} \alpha_{l_1}^G \mathbf{a}\left(\vartheta_{l_1}^{G_{\mathrm{r}}}, \psi_{l_1}^{G_{\mathrm{r}}}\right) \mathbf{u}_b\left(\vartheta_{l_1}^{G_{\mathrm{t}}}, \psi_{l_1}^{G_{\mathrm{t}}}\right)^T,
\end{aligned}\end{small}
\end{equation}
where $L_G$ represents the number of paths between the BS and the RIS, $\alpha_{l_1}^G, \vartheta_{l_1}^{G_{\mathrm{r}}} \left(\psi_{l_1}^{G_{\mathrm{r}}}\right)$, and $\vartheta_{l_1}^{G_{\mathrm{t}}}$ $\left(\psi_{l_1}^{G_{\mathrm{t}}}\right)$ represent the complex gain, the azimuth (elevation) angle at the RIS, and the azimuth (elevation) angle at BS $b$ for the $l_1$-th path. $\mathbf{u}_b\left(\vartheta, \psi\right) \in \mathbb{C}^{N_t \times 1}$ and 
$\mathbf{a}\left(\vartheta, \psi\right) \in \mathbb{C}^{N \times 1}$ represent the normalized array steering vector associated to BS $b$ and the RIS, respectively. For a typical $N_1 \times N_2 \left(N = N_1 \times N_2 \right)$ uniform planar array, $\mathbf{a}\left(\vartheta, \psi\right)$ can be represented by \cite{channelmodel1}
\begin{equation}
\begin{small}
\begin{aligned}
\mathbf{a}(\vartheta, \psi)=\frac{1}{\sqrt{N}}\left[e^{-j 2 \pi d \cos (\psi) \mathbf{n}_1 / \lambda}\right] \otimes\left[e^{-j 2 \pi d \sin (\psi) \cos (\theta) \mathbf{n}_2 / \lambda}\right],
\end{aligned}\end{small}
\end{equation}
where $\mathbf{n}_1 = [0,1,...,N_1-1]^T$ and $\mathbf{n}_2 = [0,1,...,N_2-1]^T$, $\lambda$ is the carrier wavelength, and $d$ is the antenna spacing which is typically set to $d = \lambda / 2$.

Similarly, the channel matrix  between RIS to UE $k$, denoted by $\mathbf{v}_{k}^{H}$ can be represented by

\begin{equation}
\begin{small}
\begin{aligned}
\mathbf{v}_{k}^{H}=\sqrt{\frac{N}{L_{k}}} \sum_{l_2=1}^{L_{k}} \alpha_{l_2}^{k} \mathbf{a}\left(\vartheta_{l_2}^{k}, \psi_{l_2}^{k}\right)^T,
\end{aligned}\end{small}
\end{equation}
where $L_k$ represents the number of paths between the RIS and the $k$-th user; $\alpha_{l_2}^G, \vartheta_{l_2}^{G_{\mathrm{r}}}$, and $\psi_{l_2}^{G_{\mathrm{r}}}$ represent the complex gain, the azimuth angle, and the elevation angle at the RIS for the $l_2$-th path. 

It is assumed that the channels between  BSs and  UEs are completely obstructed. It is worth noting that the model can be easily extended to scenarios involving direct BS-UE links. Given the influence of cascaded channel gain quality on the received signal quality from the BS, this study focuses on the estimation of cascaded channel gains.

Consequently,  the signal received by UE $k$  from BSs is
% \begin{equation}\vspace{-0.3em}
% \begin{aligned}
% y_{b, k} &=\mathbf{v}_{k}^H\boldsymbol{\Phi} \mathbf{G}_{b}\mathbf{f}_{b, k} {s}_{k}+{n}_{k} =\psi{\mathbf{h}}_{b, k}\mathbf{f}_{b, k} {s}_{k}+{n}_{k}. 
% \label{received signal}\vspace{-0.3em}
% \end{aligned}
% \end{equation}
\begin{equation}\vspace{-0.5em}
\begin{aligned}
y_{k} &=\sum_{b=1}^B \left\{ \mathbf{v}_{k}^H \boldsymbol{\Phi} \mathbf{G}_{b} \mathbf{f}_{b, k} \right\} {s}_{k}+{n}_{k} =\phi{\mathbf{h}}_{k}\mathbf{F}_{k} {s}_{k}+{n}_{k},
\label{received signal}
\end{aligned}\vspace{-0.3em}
\end{equation}
where ${s}_{k}$ represents the pilot signal, $\phi=$diag$(\boldsymbol{\Phi})\in \mathbb{C}^{1 \times N}$, ${\mathbf{h}}_{k}=$diag$(\mathbf{v}_{k}^H)\mathbf{G}\in \mathbb{C}^{N \times \left(N_t \times B\right)}, \mathbf{G} = [\mathbf{G}_1,...,\mathbf{G}_b,...,\mathbf{G}_B] \in \mathbb{C}^{N \times \left(N_t \times B\right)}$ represents the cascaded channel between BSs and UE $k$. Note that during the channel estimation process, both the phase shift matrix $\boldsymbol{\Phi}$ of RIS and the beamforming vector $\mathbf{F}_{k}$ at BSs are predetermined and known to both communication parties like pilot signals. Thus, the task of downlink channel estimation is to estimate the channel matrix ${\mathbf{h}}_{k}$ given $\boldsymbol{\Phi}$ and $\mathbf{F}_{k}$.

To address the computational challenge, a federated learning-based approach is introduced for downlink cascaded channel estimation. The following two subsections introduce the deep neural network (DNN) model utilized at the UE. Building upon this, a transfer-learning based model is presented.

\vspace{-0.5em}\subsection{ Basic DNNs Model for Local Channel Estimation}
At the UE, a DNN is deployed for downlink cascaded channel estimation, establishing a non-linear mapping between received pilots and cascaded channel parameters. This complex mapping is mathematically represented by
\begin{equation}\vspace{-0.3em}
\begin{aligned}
\hat{\mathbf{h}}_{b, k} = {f}_{\boldsymbol{\omega}}(y_{b, k}),
\label{mapping}\vspace{-0.3em}
\end{aligned}
\end{equation}
where ${f}_{\boldsymbol{\omega}}(\cdot)$ stands for the non-linear mapping function governed by the weight parameters $\boldsymbol{\omega}$, $\hat{\mathbf{h}}_{b, k}$ is the estimated channel matrix from the $b$-th BS to $k$-th UE. To effectively train the DNN, the UE accumulates a sufficient volume of training data beforehand. The training dataset is defined by $\left\{\mathbf{y}_{b, k}, {\mathbf{h}}_{b, k}\right\}_{D_{b, k}},\forall b, k$, where $\mathbf{h}_{b, k}$ is the target channel matrix from the $b$-th BS to $k$-th UE, and $D_{b, k}$ represents the size of training dataset for the $k$-th UE served by the $b$-th BS. The loss function can be represented by 
\begin{equation}\vspace{-0.3em}
\begin{aligned}
\mathcal{L}(\boldsymbol{\omega})=\frac{1}{D_{b, k}} \sum_{D_{b, k}}\left\|\hat{\mathbf{h}}_{b, k}-\mathbf{h}_{b, k}\right\|_2^2,
\label{Loss}\vspace{-0.3em}
\end{aligned}
\end{equation}

The   goal of training the DNN is to minimize the loss function   by   optimizing the weight parameters ${\boldsymbol{\omega}}$, i.e.,
% The aim of training the DNN is to minimize the above loss function by optimizing the weights {\boldsymbol{\omega}}{\boldsymbol{\omega}}, i.e.,
\begin{equation}\vspace{-0.3em}
\begin{aligned}
\min _{\boldsymbol{\omega}} \mathcal{L}(\boldsymbol{\omega})=\frac{1}{D_{b, k}} \sum_{D_{b, k}}\left\|\boldsymbol{f}_{\boldsymbol{\omega}}\left(y_{b, k}\right)-\mathbf{h}_{b, k}\right\|_2^2 .
\label{Loss_min}\vspace{-0.3em}
\end{aligned}
\end{equation}

Upon defining the objective function, the DNN undergoes iterative optimization using the given training dataset. In each iteration $t$, the weights $\boldsymbol{\omega}$ are updated using gradient descent, i.e.,
\begin{equation}\vspace{-0.5em}
\begin{aligned}
\boldsymbol{\omega}_{t+1}=\boldsymbol{\omega}_t-\eta_t \mathbf{g}\left(\boldsymbol{\omega}_t\right),
\label{Iteration}\vspace{-0.3em}
\end{aligned}
\end{equation}
where $\boldsymbol{\omega}_{t}$ and $\boldsymbol{\omega}_{t+1}$ are the weights in the $t$-th   and $(t+1)$-th iterations for DNN, respectively. Additionally, $\mathbf{g}\left(\boldsymbol{\omega}_t\right)$ is the gradient vector for $\boldsymbol{\omega}_{t}$, and $\eta_t$ is the learning rate. After training the DNN, the UE can estimate the channel directly based on the trained DNN.

However, the requirement of sending user data to a centralized node, such as a BS, for unified training incurs substantial overhead and raises significant privacy concerns. To address this issue, we utilize a distributed learning paradigm known as FL, which enables decentralized training.

\vspace{-0.5em}\subsection{FL-based Channel Estimation Network Training}
Compared with DNN-based methods, the proposed FL-based method reduces communication overhead and enhances privacy, as FL only requires the exchange of model parameters. This leads to a significant reduction in communication costs. 
The FL integrates  DNN and can be fundamentally divided into several procedures:
\begin{itemize}
    \item \emph{FL user group selection}: The central node (CN) identifies a subset of UEs, denoted by $\widehat{\boldsymbol{M}} \subset \left\{1, 2, ..., M\right\}$ to participate in the FL process. We refer to the UEs in $\widehat{\boldsymbol{M}}$ as the active UEs.

    \item \emph{Model broadcast}: The CN broadcasts the current $t$-th global model $\boldsymbol{\omega}_{t}^{g}$ to the active UEs.

    \item \emph{Local gradient calculation}: Each active UE calculates its local gradient by utilizing its dataset. The gradient  is  achieved as  $\mathbf{g}\left(\boldsymbol{\omega}_{m,t}^{l}\right) \triangleq \nabla \mathcal{L}(\boldsymbol{\omega}_{t}^{g}, D_{m}),$ where $\nabla \mathcal{L}(\boldsymbol{\omega}_{t}^{g}, D_{m})$ is the gradient of $\mathcal{L}$ at $\boldsymbol{\omega}_t = \boldsymbol{\omega}_t^{g}$.

    \item \emph{Model aggregation}: Active UEs transmit their computed local gradients $\mathbf{g}\left(\boldsymbol{\omega}_{m,t}^{l}\right)$ to the CN via wireless channels. Subsequently, the CN aims to compute a weighted aggregate of these local gradients $\left\{ \mathbf{g}\left(\boldsymbol{\omega}_{m,t}^{l}\right): m \in \widehat{\boldsymbol{M}}_t\right\}$, which is used to update the global model. Adopting the prevalent FL configuration from \cite{FL-model}, the weight of each local gradient vector is normalized. The global gradient vector can be thus represented as 
    \begin{equation}\vspace{-0.3em}
    \begin{aligned}
    \mathbf{r}_t \triangleq \sum_{m \in \widehat{\boldsymbol{M}}_t} \left\{ \frac{ \alpha_{m}}{\sum_{i \in \widehat{\boldsymbol{M}}_t} \alpha_{i}} \mathbf{g}\left(\boldsymbol{\omega}_{m,t}^{l}\right)\right\},\vspace{-0.3em}
    \label{r_t}
    \end{aligned}
    \end{equation}
    where $\alpha_{m}$ represents the corresponding weight attributed to UE $m$'s contribution in the update of the global model's parameters, such as distance weight. Therefore, (\ref{Iteration}) can be written as 
    \begin{equation}\vspace{-0.3em}
    \begin{aligned}
    \boldsymbol{\omega}_{t+1}=\boldsymbol{\omega}_t-\eta_t{\mathbf{r}}_t.\vspace{-0.3em}
    \end{aligned}
    \end{equation}
\end{itemize}

\section{Coalition Formation for FL Enabled Channel Estimation}
In this section, we design a coalition formation-guided FL framework for channel estimation by integrating the previously introduced DNN and FL approaches. First, each local user trains the DNN using the data that they have collected. Next, to reduce communication overhead and improve channel estimation accuracy, we introduce a coalition formation-guided FL (CFFL) architecture, where a DRL method is employed to intelligently guide the coalition formation. Finally, to enhance the efficiency of the framework, we further integrate transfer learning into the CFFL method.

\vspace{-0.5em}\subsection{DNN for the Downlink Channel Estimation (DCE)}
To augment the volume of data available to individual UEs and enhance estimation accuracy,  we present a strategy wherein geographically proximate neighboring UEs, within a certain radius from the active UE, contribute a portion of their data to train the DNN network of the active UE. This approach not only ensures an adequate supply of training samples but also minimizes communication overheads compared with conventional distributed ML paradigms. Therefore, a higher learning efficiency and lower transmission overhead can be achieved.

A deep neural network (DNN) architecture is adopted, comprising three convolutional layers followed by a linear layer. Each convolutional layer includes a series of operations: a convolutional operation, a batch normalization operation, and an activation function, respectively

Considering that both received pilot signal $y_{b, k}$ and the cascaded channel ${\mathbf{h}}_{b, k}$ are complex, the DNN inputs and outputs are transformed to real-valued representations. Specifically, $y_{b, k}$ is reshaped into a $Q_1 \times Q_2$ 2-dimensional matrix, with its real and imaginary components serving as dual input feature maps for the first convolutional layer. The DNN output is structured as a $2N \times M$-element vector, where the first  $N \times M$ entries correspond to the real part of ${\mathbf{h}}_{b, k}$, and the remaining $N \times M$ entries represent its imaginary counterpart.

Without loss of generality, we assume that   UE $k$ is surrounded by a total of $Z$ neighboring UEs. The aim of training the DNN is to minimize the sum of the loss functions by optimizing the weights ${\boldsymbol{\omega}}$, i.e.,
\begin{equation}\vspace{-0.3em}
\begin{small}
\begin{aligned}
\min _{\boldsymbol{\omega}} \mathcal{L}(\boldsymbol{\omega})&=\frac{1}{D_{k}} \sum_{D_{k}}\left\|\boldsymbol{f}_{\boldsymbol{\omega}}\left(y_{k}\right)-\mathbf{h}_{k}\right\|_2^2 \\ &+ \sum_{z=1}^{Z} \left\{ \frac{1}{D_{z}} \sum_{D_{z}}\left\|\boldsymbol{f}_{\boldsymbol{\omega}}\left(y_{z}\right)-\mathbf{h}_{z}\right\|_2^2 \right\},
\label{Loss_sum_min}\vspace{-0.3em}
\end{aligned}\end{small}
\end{equation}
where $D_z \geq a D_{k}$, $a$ is an empirical value, generally set to $0.1$\cite{TLcite}.

\vspace{-0.5em}\subsection{Transfer Learning-based Channel Estimation}
The FL process involves the iterative training of local models followed by the aggregation of a global model, which can result in prolonged training times. To accelerate model convergence, while maintaining high accuracy in channel estimation, two types of transfer learning strategies are documented. The fundamental concept is that the DNN network weights of a central node are not trained from scratch; instead, they are initialized based on the trained DNN network weights of neighboring central nodes. This initialization incorporates a weighted aggregation of the weights from neighboring nodes, considering both distance proximity and the similarity of the received reference signal power (RSRP)\cite{RSRP} to these nodes. This approach facilitates the determination of the central node's initial DNN network parameters; thereby, accelerating subsequent DNN parameter iterations. The details of the two types of transfer learning strategies are presented in the following sections.

The first strategy is distance (Dis)-similar transfer learning. Given that users close to one another are likely to exhibit similar environmental conditions and channel characteristics, transferring federated channel estimation models across proximate spatial locations not only accelerates training but also maintains a certain level of accuracy. The weight under the first transfer strategy is represented as
\begin{equation}\vspace{-0.3em}
\begin{aligned}
\alpha_{m}^{'}=\left\{\begin{array}{l}1-\frac{L_m}{L}, L_m \leq L, m=1,2, \ldots, K \\ 0, \text { otherwise }\end{array}\right.
\vspace{-0.3em}
\end{aligned}
\label{am'}
\end{equation} where $L_m$ represents the distance between UE $m$ and the CN aggregating the global model, and  $L$ is the predefined reference distance for normalization. Finally, replacing $\alpha_{m}^{'}$ with $\alpha_{m}$ in \eqref{r_t} results in the ``FL+Dis" strategy.

The second strategy of FL is based not only on Dis-similarity, but also integrates power similarity (hereinafter referred to as ``RSRP+Dis".This is motivated by the insights drawn from BS association rules prevalent in cellular networks. When the received signal power of two neighboring users closely matches the RSRP\cite{RSRP}, it suggests a strong likelihood that the two signals of neighboring users have experienced similar channel fading; hence, sharing comparable channel fadings. 

Consequently, the weight under the second transfer learning strategy can be formalized as 
\begin{small}
\begin{equation}
\alpha_{m}^{''}=\begin{cases}\begin{array}{l}
\left\{1-\frac{\left|P_m-P_{\text {center}}\right|}{2 \Delta}\right\}\alpha_{m}^{'}, P_{\text {center}}-\Delta<P_m<P_{\text {center}}+\Delta \\
0, \text { otherwise}
\end{array},\end{cases}
\label{am''}
\end{equation}
\end{small}where $P_m$ represents the power of the channel sample at UE $m$, while $P_{\text {center}}$ represents the predetermined central reference power level at the center user, $\Delta$ represents the predefined range of comparable power levels \cite{TLcite}. Finally, replacing $\alpha_{m}^{''}$ with $\alpha_{m}$ in \eqref{r_t} results in the ``FL+RSRP+Dis" strategy.

\vspace{-0.5em}\subsection{HFL Enabled Channel Estimation}
\begin{figure}[htb]
    \centering
    \includegraphics[width=0.73\linewidth]{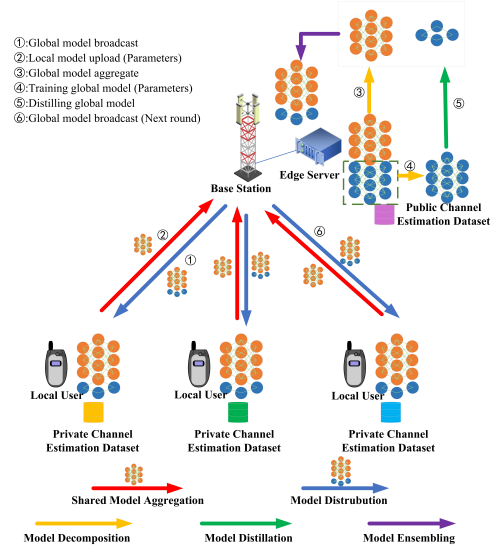}
    \caption{Aggregation process of heterogeneous federated learning model.}
    \label{Heterogeneous Federated-learning}\vspace{-1em}
\end{figure}

In the above application of the FL algorithm, the CN's and users' devices share the same neural network structure as well as data properties,  and are therefore referred to as homogeneous FL; whereas in reality, the local user generally has very limited computational and power capacity and is not capable of handling neural network training with high computational complexity and power consumption. To address that, a HFL-based approach is proposed that reduces the size of local neural network models for individual users, leading to a substantial decrease in computational costs and power consumption during the training process compared to homogeneous FL. The algorithm for HFL is described in Figure \ref{Heterogeneous Federated-learning}.

The whole channel estimation process is executed iteratively, with each round including 6 steps:
\begin{figure}[htb]
    \centering
    \includegraphics[width=0.93\linewidth]{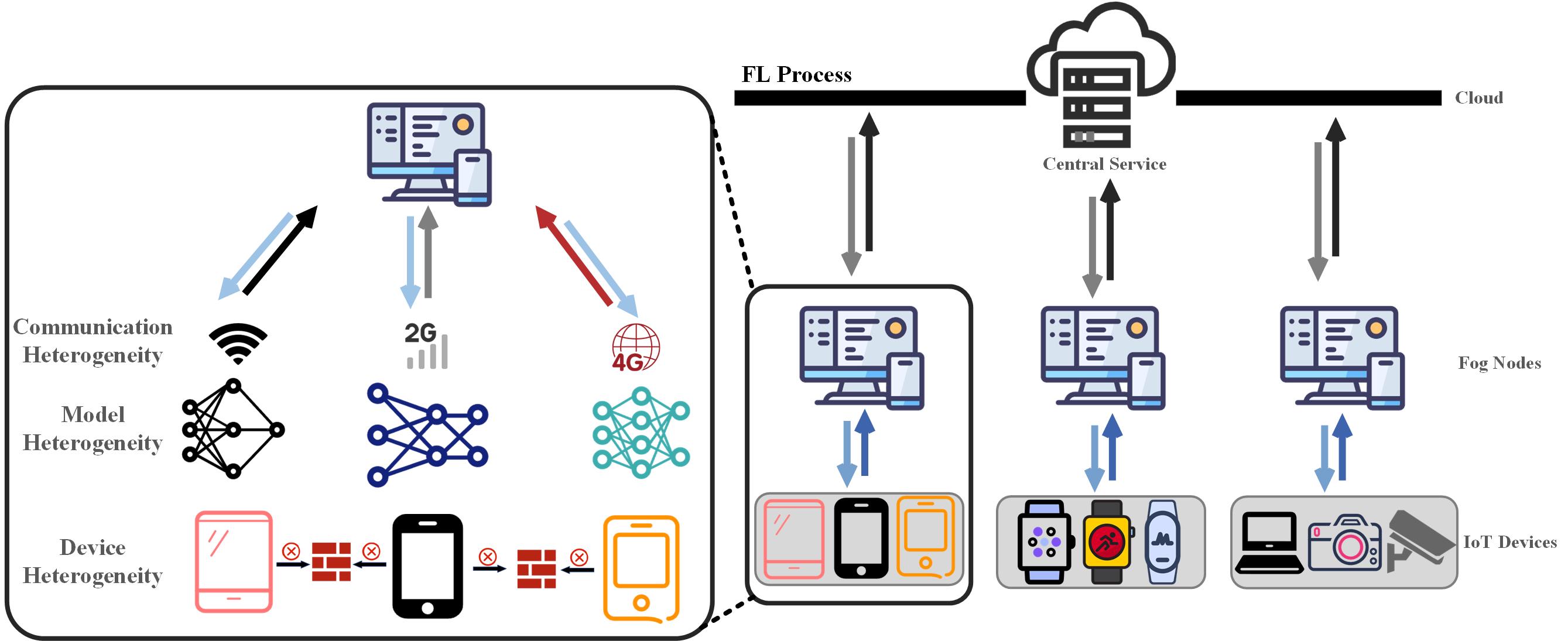}
    \caption{The concept of heterogeneous federated learning networks.}
    \label{HFL_detail}\vspace{-1em}
\end{figure}
\begin{figure}[h]
    \centering
    \includegraphics[width=0.65\linewidth]{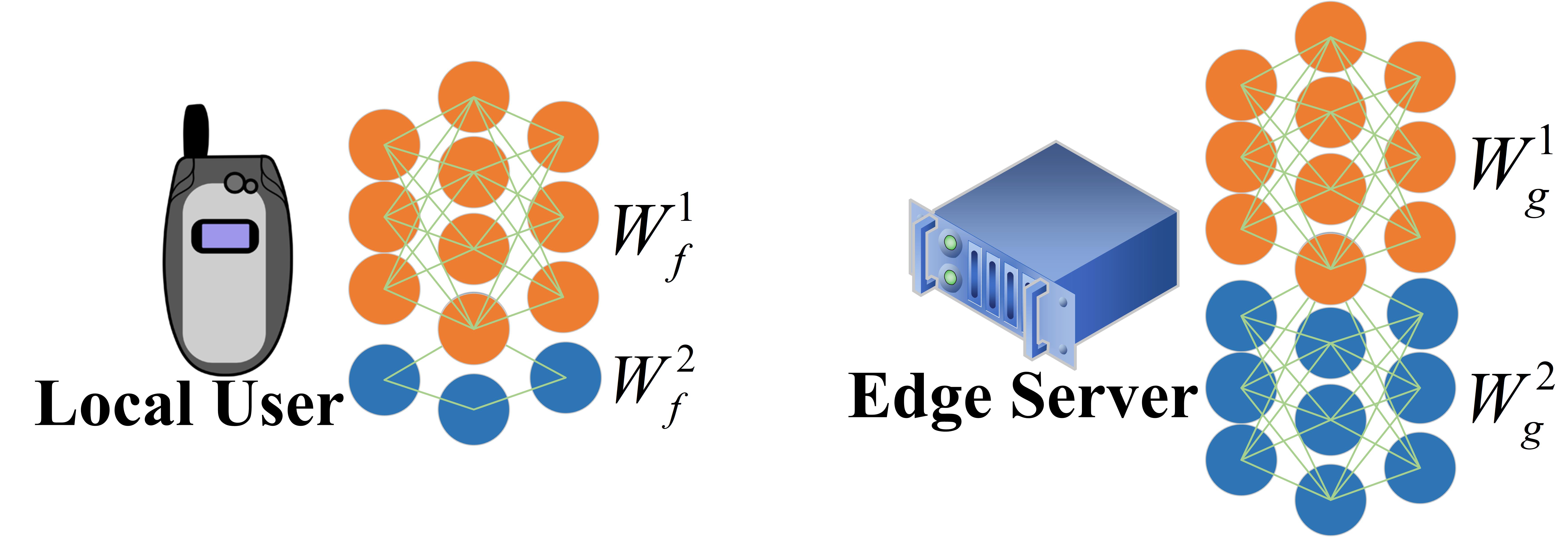}
    \caption{Composition of models.}
    \label{Model}\vspace{-0.5em}
\end{figure}
% \paragraph{The HFL Based Training}
% \newline

%{\textbf{Step 1:}}
\emph{Step 1- Global model broadcast:} A central/edge server or node leverages its superior computing capabilities to provide a powerful global model. A global model $W_{g}$ is initialized at the central server, which is larger than the users' private neural network models. Within the heterogeneous FL framework, the global model $W_{g}$ is decomposed into the shared part $W_{g}^1$ and distillation part $W_{g}^2$, as shown in Figure \ref{Model}. $W_g$ is distilled into a small neural network $W_f$ that is broadcast to and used by local users. Note that $W_f$ is composed of $W_f^1$ and $W_f^2$ as depicted in Figure \ref{Model}. The structure of $W_g^1$ is identical to that of $W_f^1$, but with different parameters.

%{\textbf{Step 2:}}
\emph{Step 2- Local model upload:} After receiving $W_f$, each local users train their individual local models to minimize MSE (i.e., update $W_f$ independently\footnote{Note that though we assume all users share the same neural network size,  it can be easily extended to situations where users have different model size.}) by utilizing their respective data set collected by themselves. Finally, the local user $l$’s private local model is obtained, denoted by $W_l$, which can be decomposed into the shared part $W_{l}^1$ and distillation part $W_{l}^2$.

%{\textbf{Step 3:}}
\emph{Step 3- Shared part aggregating of the global model:} After user $l$ completes training its local model, the user proceeds to transmit the parameters of the shared part $W_{l}^1$ through the dedicated channel. 
\begin{figure}[h]
    \centering
    \includegraphics[width=0.82\linewidth]{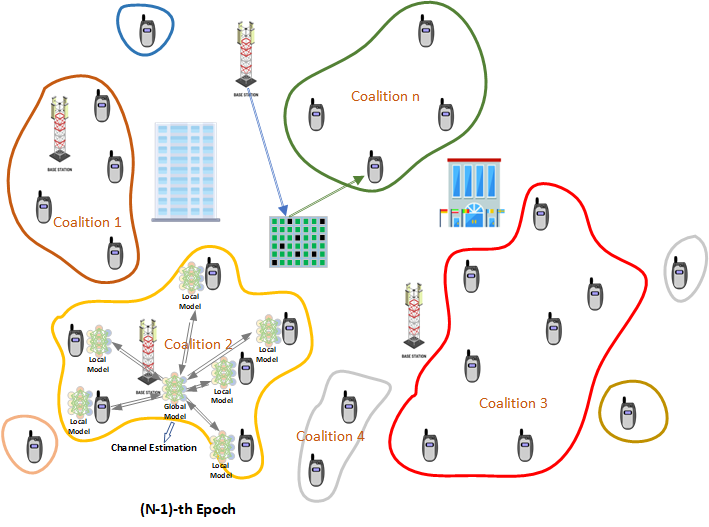}
    \caption{Coalition formation diagram for channel estimation in RIS-assisted cell-free MIMO networks.}
\label{System_Model_for_coalition}\vspace{-1.5em}
\end{figure}
The central server first aggregates the parameters of the shared part based on the average criterion and generates the shared part parameters of the global model, that is 
\begin{equation}\vspace{-0.3em}
\begin{aligned}
W_g^1=(\sum_{l=1}^L \beta_l W_l^1) / L,
\vspace{-0.3em}
\end{aligned}
\end{equation}where $L$ is the number of users, $\beta_l$ is the weight of $l$-th model parameter in the global model aggregation.

% {\textbf{Step 4:}}

\emph{Step 4- Distillation part training of the global model: } The central server proceeds to train the distillation part $W_{g}^2$ through part of the data set. Note that to reduce the user's downlink data privacy, we use the data collected by the edge server. Specifically, we consider the reciprocity of uplink and downlink channels\cite{channelreciprocity}. During the uplink transmission phase, the data uploaded by users to BS are stored as training data samples for this phase. By estimating the downlink channel and leveraging channel reciprocity, we obtain the uplink channel information.

After Steps $3$ and $4$, a global model is generated. Different from traditional homogenous FL approach, we first distill the global model before broadcasting it to the local users.

%{\textbf{Step 5:}}
\emph{Step 5-  Global model distilling}: Specifically,   local user model $W_f$ is obtained by distilling $W_g$. The network structure of $W_f$   is identical to that of the user's model. $W_f^{1}$ shares the same model parameters as $W_g^{1}$, while  the parameters of the distillation part  $W_{l}^2$ is initialized independently. In the training process, the data set is utilized, and only the distillation part, i.e., $W_f^{2}$ is iteratively updated till convergence is achieved or the maximum epoch is reached. Then, model $W_f$, which has the same network structure as the user's local model, is finally obtained.

% {\textbf{Step 6:}}
\emph{Step 6- Global model broadcast:} Next round of  HFL starts. The central server broadcasts $W_f$ to the users for  the subsequent round of channel estimation training. Note that after training, the server uses the global model $W_g$, while the local users utilize the distilled models $W_f$ for the HFL process.

\begin{remark}
Under the HFL framework,  the size of local neural network models for individual users is reduced, leading to a substantial decrease in computational and power costs during the training process, compared to homogeneous FL. While HFL results in a slight reduction in accuracy, it significantly lowers computational costs (measured in flops) and complexity. Additionally, under the homogeneous FL framework,  only a portion of the model parameters is transmitted during the training process. \begin{figure}[h]
    \centering
    \includegraphics[width=1\linewidth]{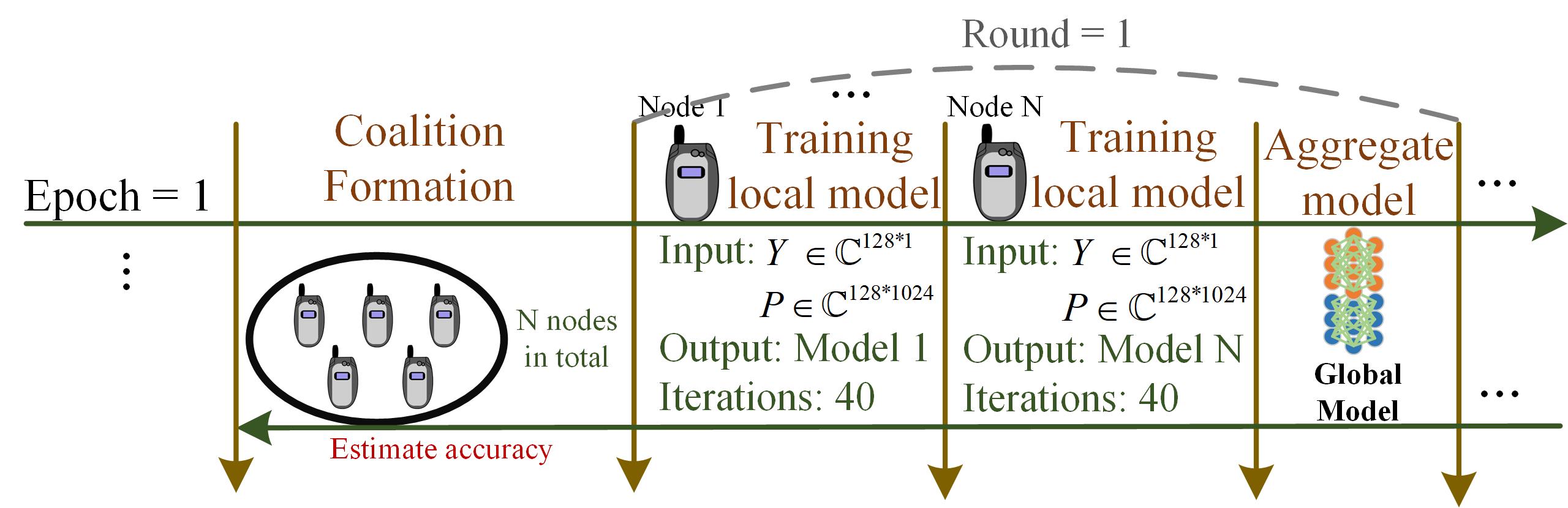}
    \caption{CFFL for channel estimation in RIS-assisted cell-free MIMO networks.} %(Scenario parameter settings: one BS antenna: 4*4=16 antennas, RIS: 8*8=64 units, number of channels: 64*16.)
    \label{System_Model_for_coalition}\vspace{-1em}
\end{figure}This approach reduces the quantity of data exchanged; thereby, enhancing privacy protection for individual users. 
\end{remark}

\vspace{-1em}\subsection{Coalition Formation-enabled FL Method}
To enhance channel estimation accuracy in FL frameworks, whether homogeneous or heterogeneous, optimizing the user grouping strategy is an effective approach. This optimization facilitates the construction of a high-quality aggregation neural network; thus, improving channel estimation accuracy.  In this context, a coalition is a FL user group. The FL user grouping problem is converted into a coalition formation  problem. Accordingly,  a CFFL method is documented to improve the accuracy of the HFL approach.The details  are described as follows. 
First, the initial coalition combinations are determined using a reinforcement learning algorithm. Each user employs its local dataset to perform DNN-based channel estimation, generating a locally trained model. The model parameters from local users are then transmitted to the central node of each coalition via the uplink channels. Subsequently, the central node performs federated learning model aggregation and distributes the aggregated and distilled model to each user within the coalition. The global utility of the current coalition combination is evaluated by averaging the normalized mean square error (NMSE) across all users within the coalition. This NMSE value serves as a reward for the reinforcement learning algorithm, which updates the coalition combinations iteratively. This process is repeated until convergence. In what follows, we present two reinforcement learning approaches of coalition formation, namely i) DQN-enabled coalition formation, and ii) Qmix-enabled coalition formation.

\subsubsection{DQN-enabled Coalition Formation}
% In the sequel, we propose a coalition formation-enabled federation learning framework for optimizing the FL user group.  This framework utilizes
In this section, DQN is utilized  to solve the  FL coalition formation problem. The flow diagram, which is presented in Figure \ref{System_Model_for_coalition} illustrates the initial round of the first epoch in CFFL. Following this, a total of $T$ epochs and each epoches consists of $E$ rounds are executed. Then we introduce the state, action, and reward of the DQN in the CFFL framework as follows.
\begin{figure}[h]
    \centering
    \includegraphics[width=0.82\linewidth]{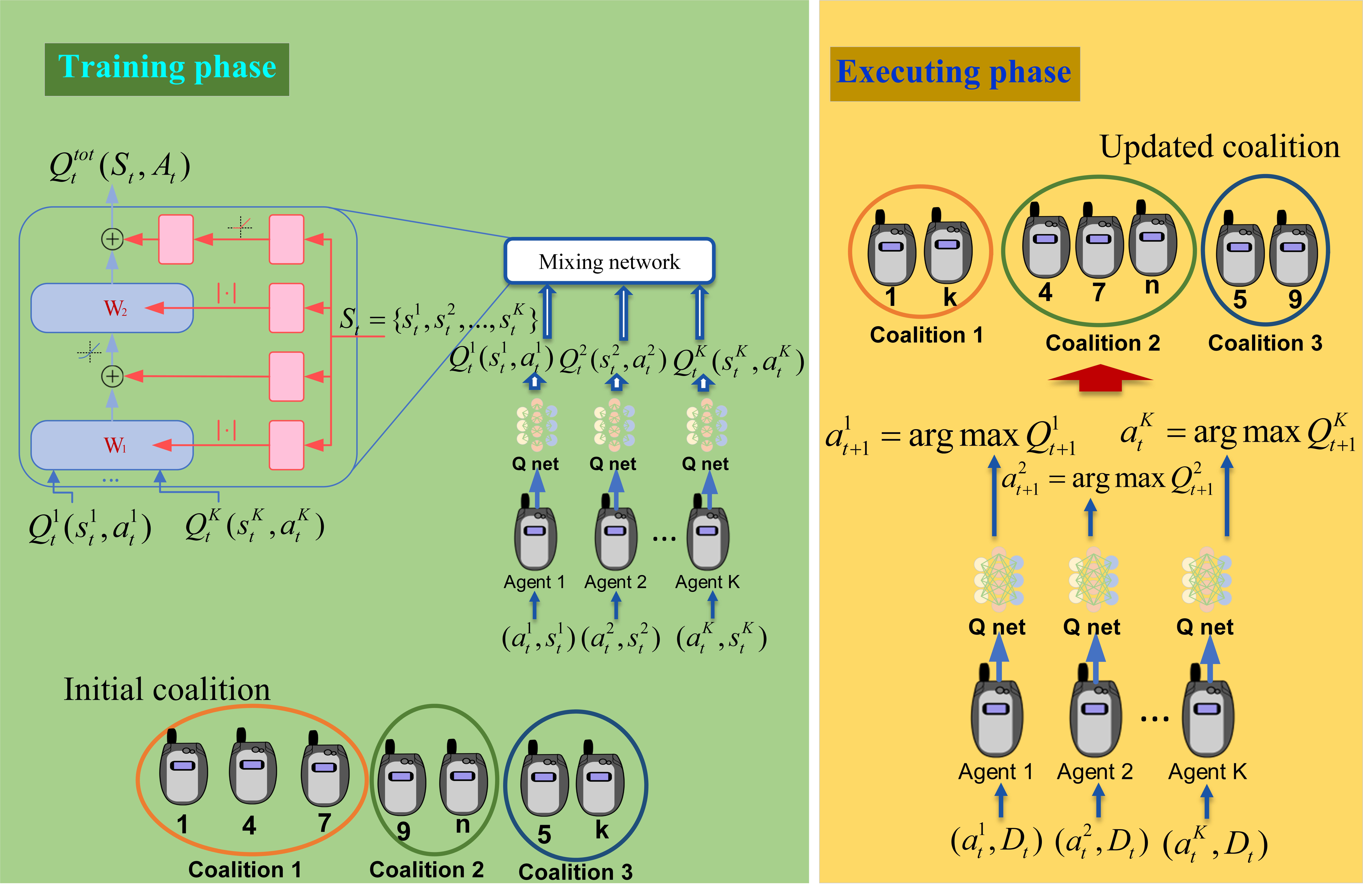}
    \caption{Qmix-enabled coalitions formation.}
    \label{Qmix enables coalition formation}\vspace{-1em}
\end{figure}
\begin{itemize}
    \item $State:$ $S_t=\left\{C_t, D_t\right\}$ is defined as the state information of the coalition combination at $t$-th epoch, in which $C_t=\left\{c_{1,t},...,c_{K,t}\right\} \in \mathbb{R}^{1 \times K}$ (with $ c_{k,t}={0,1,.., J}$) is the current coalition choice of $K$ users, $c_{k,t}=0$ represents the $k$-th user who does not join any coalition and trains on its own, $J$ denotes the maximum number of coalitions. $D_t \in \mathbb{R}^{1 \times J}$ is the set of number of users in $J$ coalitions.

    \item $Action:$ $A_t = \left\{C_t\right\}$ is the coalition choice of all users.

    \item $Reward:$ $R_t = 1 - \overline{e}$, in which $\overline{e}$ represents the average channel estimation accuracy of all users after one epoch of CFFL.

\end{itemize}

\subsubsection{Qmix-enabled CFFL}
DQN is a centralized algorithm that constructs a large joint action-value mapping function from global observations. However, it may lead to a rapid increase in complexity as the number of users grows, while failing to satisfy privacy-preserving and communication-efficient requirements. Alternatively, distributed frameworks can be applied, wherein each agent has its local observation and makes decisions independently. Qmix is a distributed reinforcement learning algorithm, where each agent has its individual value function. These individual value functions are then combined into a global value function for the entire team through a centralized mixing network. Correspondingly, the distributed framework facilitates autonomy and collaboration among the users in the coalition, where each user can make decisions based on local information, while also taking into account the interests of the whole coalition with the help of the mixing network.

In Qmix architecture, each user can only achieve the observation of which coalition it joins and has no information on the decisions of other users. This design eliminates the need for information exchange, which consequently reduce communication costs associated with tracking other users' coalition choices.
In the training process, the coalition is initialized. Based on their current action and the observed state $S$, including $\left\{C, D\right\}$, each user calculates the Q-value, as shown in Figure \ref{Qmix enables coalition formation}. Subsequently, the Q-values obtained by each user along with all the information about the observed environment will be uniformly input into the mixing network. The network is trained   to generate the global Q-value as well as the Q-value weights for each user. Each user's Q network is iteratively  updated to maximize    the global Q value until convergence or the end of training.

During the execution phase, users don't need to go through the mixing network for centralized execution; they can execute actions through their own Q network by using their current action and the information concerning the number of users in the coalition.

\subsubsection{FL User Coalition Formation Game}
The traditional coalition formation game divides the set of players into several disjoint sets through the cooperation between users, and the players of the game pursue the maximization of their own interests. In this section, the participating UEs in channel estimation are  game players, forming distinct coalitions, where members within each coalition engage in model aggregation through federated learning. According to the previous analysis, we configure the utility of the entire network as the aggregate sum of utilities across all coalitions. The coalition game model can be represented as 
\begin{equation}
\mathcal{G}=\left\{\mathcal{K}, U(S), S\right\},
\end{equation}
where $\mathcal{K}$ is the set of UEs participating in the game for channel estimation, $S=\left\{S_1, ..., S_J\right\}$ represents the coalition structure formed among these UEs. $c_{k,t}=j$ means that, at $t$-th epoch, UE $k$ belongs to coalition $S_j$, and $U(S_j)$ is the utility of coalition $j$.

The utility all coalitions is formulated as
\begin{equation}\begin{small}
\begin{aligned}
U=\sum_{S_j \in S}U\left(S_j\right),\vspace{-0.3em}
\end{aligned}\end{small}\end{equation}
where $U\left(S_j\right)=A-\sum_{p=1}^P e_{j,p}$ is the utility of coalition $j$, $A$ is a positive constant introduced to ensure non-negativity of utilities, and $e_{j,p}$ represents the normalized mean squared error between the estimated channel for UE $p$ within coalition $j$ and the true   channel. The utility assigned to the $p$-th user in the $j$-th coalition, denoted by $u\left(a_{j, p}\right)$, is given  by
\begin{equation}\vspace{-0.3em}
\begin{aligned}
u\left(a_{j, p}\right)=\frac{\left|a_{j, p}\right|}{\sum_{q \in S_j}\left|a_{j, q}\right|} U\left(S_j\right),
\end{aligned}
\end{equation}
 where  $\left|a_{j, p}\right|$ represents the magnitude of data volume contributed by UE $a_{j, p}$.

\subsubsection{Analysis of Nash Equilibrium (NE)}
The design of a preference order that ensures a stable coalition structure while maximizing the global utility is crucial for guiding coalition formation.  
The preference relationship of altruistic criteria between coalition $S_1$ and coalition $S_2$ is formally defined in \textbf{{Definition 1}}. 

\begin{definition}
For any UE $k \in \mathcal{K}$, and two coalitions $S_1 \subseteq S$  and $S_2 \subseteq S$, the altruitic criterion is defined as 
\begin{equation}
\begin{small}
\begin{aligned}
\mathcal{S}_1 \succ_j \mathcal{S}_2 \Leftrightarrow & u_j\left(\mathcal{S}_1\right)+\sum_{k \in \mathcal{S}_1 \backslash\{j\}} u_k\left(\mathcal{S}_1\right)+\sum_{k \in \mathcal{S}_2 \backslash\{j\}} u_k\left(\mathcal{S}_2 \backslash\{j\}\right) \\
> & u_j\left(\mathcal{S}_2\right)+\sum_{k \in \mathcal{S}_1 \backslash\{j\}} u_k\left(\mathcal{S}_1 \backslash\{j\}\right)+\sum_{k \in \mathcal{S}_2 \backslash\{j\}} u_k\left(\mathcal{S}_2\right).
\end{aligned}
\end{small}\vspace{-0.3em}
\end{equation}
\end{definition} 

It means that  when UE $k$ joins a new coalition, the utilities of both the newly formed coalition, and the original coalition to which UE $k$   previously belonged, are improved. 

% According to the above analysis, the utility of UEs is related to whether the UEs in the new coalition participate in channel estimation or not. 
\begin{theorem}
    There is at least one stable coalition structure under the altruitic criterion, and the optimal deconstruction of the utility maximization problem of the whole network is a stable coalition structure.
\end{theorem}
\emph{Proof}: Refer to Appendix A.   $\blacksquare$

\vspace{-0.5em}\section{Numerical Results and Performance Analyses}
\begin{figure}[t]
    \centering
    \includegraphics[width=0.85\linewidth]{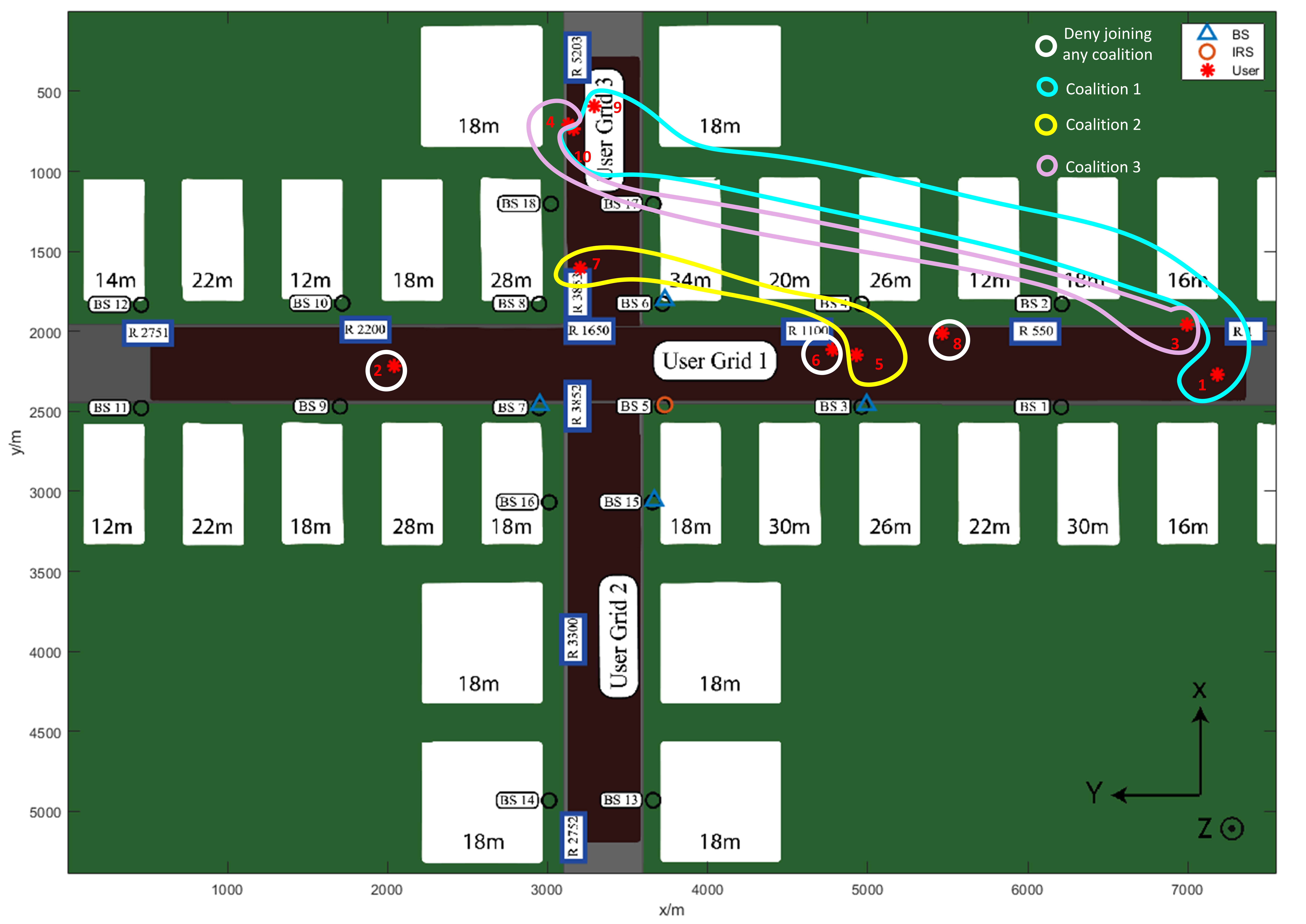}
    \caption{Sparse distribution scenario: two-dimensional geographical distribution of 10  users.}
    \label{10UE}\vspace{-1em}
\end{figure}

In the simulation, we consider a scenario comprising 4 BSs, 1 RIS, and 10 UEs. Each BS is equipped with $N_t = 16= $ $4\times4$ antennas, while the RIS consists of $N = 64 =8\times8 $ elements. All channel data utilized in this analysis were extracted from the DeepMIMO dataset \cite{DeepMIMO}. The two-dimensional geographical distribution of the 10 UEs is shown in Figure \ref{10UE}. The white box represents the distributed buildings along the streets. The main street (the horizontal one) is 600 m long and 40 m wide, the cross street (the vertical one) is 440 m long and 40 m wide. The two streets have buildings on both sides, the height of every building is marked in the figure.
\begin{figure}[t]
    \centering
    \includegraphics[width=0.82\linewidth]{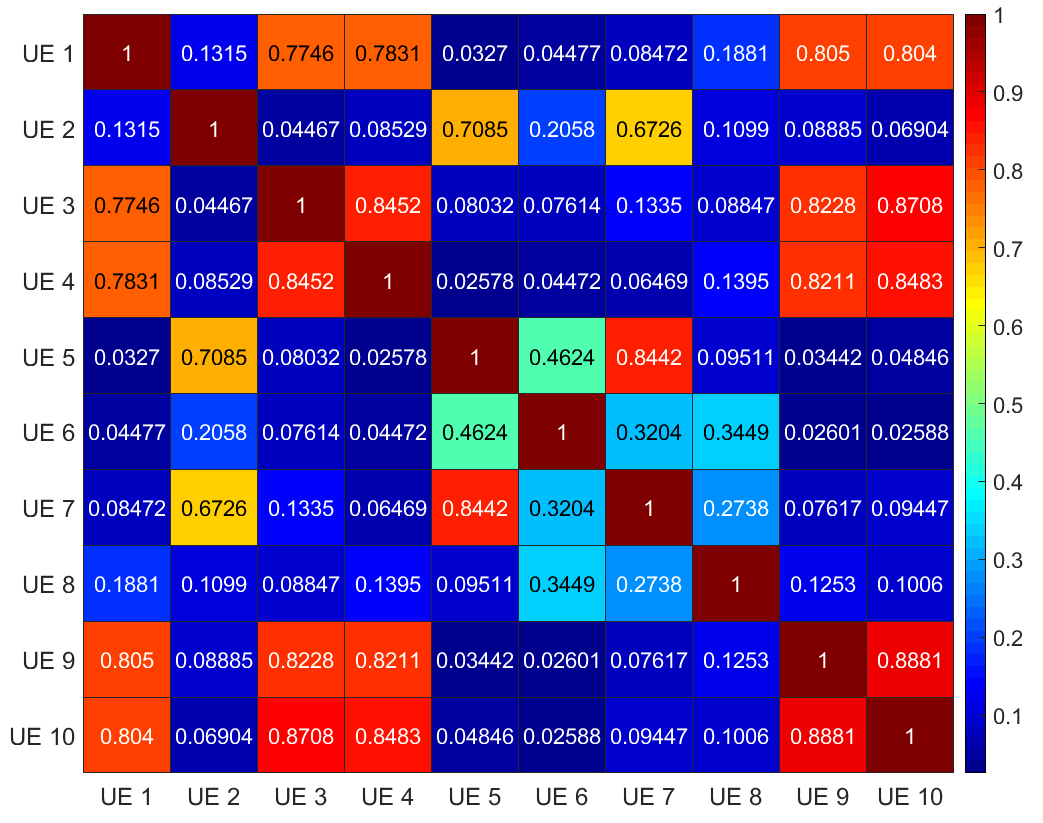}
    \caption{Channel correlation coefficient with 10 sparse users.}
    \label{10_cor}\vspace{-0.5em}
\end{figure}

Along the main street, all the buildings have bases of the same dimensions, 30 m $\times$ 60 m and the buildings have bases of 60 m $\times$ 60 m in the second street. In the HFL framework, the model of the center node has 5 convolutional layers as well as 1 fully connected layer, while the model of the edge nodes has only 3 convolutional layers and 1 fully connected layer. 
For the shake of fairness, the pilot signal length used for MMSE is set to 1024, whereas the proposed algorithm employs a pilot signal length of 128. Some of the parameters of this section are given in Table \ref{NNparameters}.

\begin{table}[h]\vspace{-0.5em}
\caption{Parameter Settings}\label{NNparameters}
\centering 
\resizebox{0.48\textwidth}{!}{
\begin{tabular}{|c|c||c|c|}

\hline
\textbf{Parameters} & \textbf{Value} & \textbf{Parameters} &  \textbf{Value} \\ \hline
Number of BSs $B$ &  4 & Number of UEs $K$ &  10     \\ \hline
Number of RISs $R$ & 1 & Number of BS antennas $N_t$ &  $4 \times 4=16$\\ \hline
Number of RIS elements $N$ & $8 \times 8=64$  & Number of samples per user &  4000\\ \hline
Batch size for training & 64 & Training test ratio &  9 \\ \hline
Learning rate & $1e^{-3}$ & Epoch&  6000\\ \hline
Maximum number of coalitions & 3 & System Bandwith  & 100MHz \\ \hline
\end{tabular}
}
\vspace{-2.0em}
\end{table}

\subsection{Impact of user channel correlation on the  resulting  formed FL user coalition}
\begin{figure}[htb]
    \centering
    \includegraphics[width=0.85\linewidth]{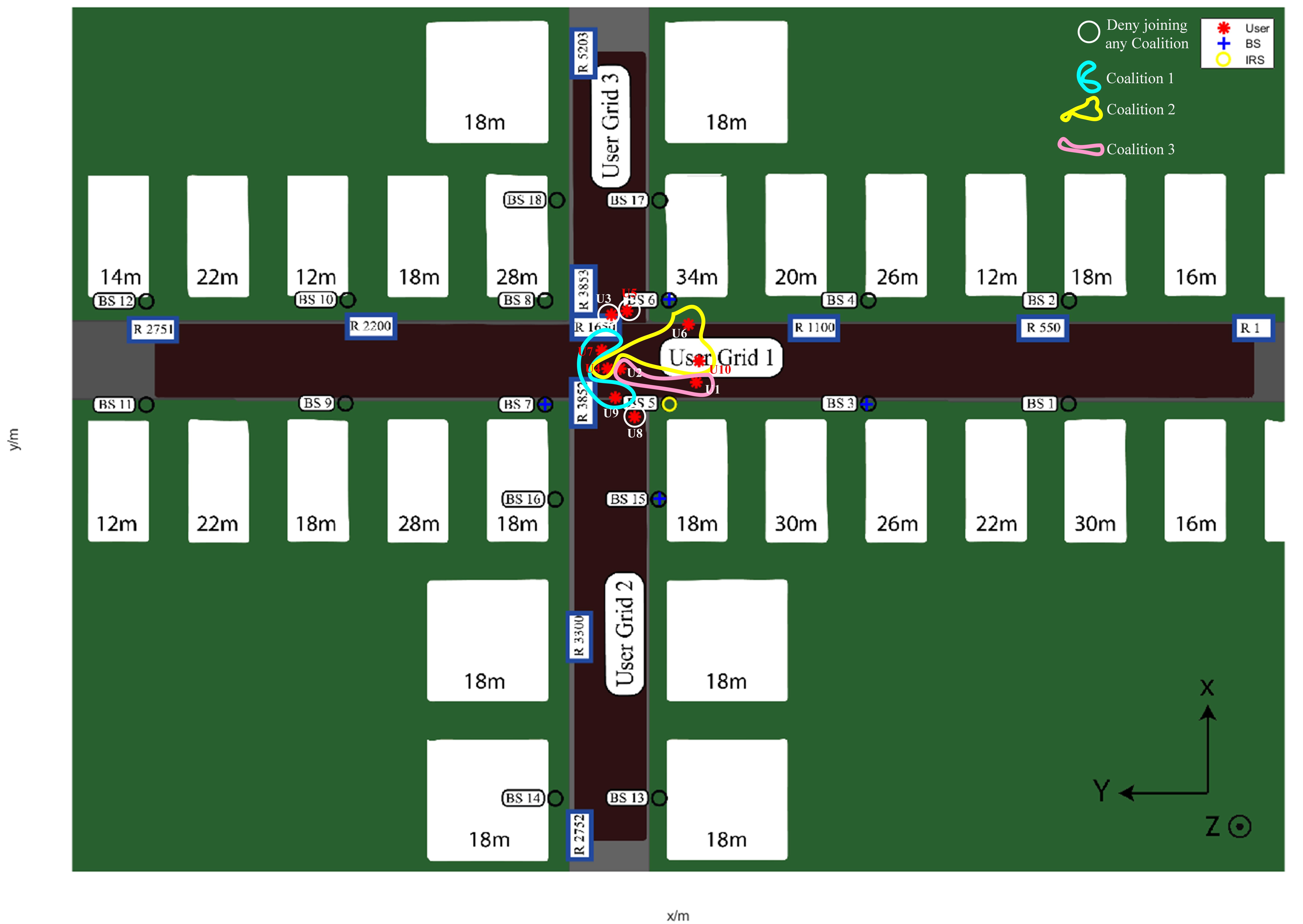}
    \caption{Dense distribution scenario: two-dimensional geographical distribution of 10  users.} 
    \label{10UE_Ro}\vspace{-1em}
\end{figure}
To verify the rationality of the resulting  coalition formation, we investigate the relationship between the formed coalition and user channel correlation.
Without loss of generality, we consider 10 UEs under two scenarios: sparse and dense distributions, as illustrated in Figure \ref{10UE} and Figure \ref{10UE_Ro}, respectively. Without loss of generality,   the maximum number of coalitions is set to 3.

For the sparse scenario, the user channel correlation coefficients are shown in Figure \ref{10_cor}. 
Of note the channel correlations among users vary, motivating the grouping of users with strong channel correlation to improve  the efficiency of federated learning. 
Applying the proposed algorithm, the resulting user actions (i.e., coalition choices) are    $A_t = [1, 0, 3, 3, 2, 0, 2, 0, 1, 1]$, where $0$ indicates that the user does not join the coalition, and  other numbers indicate which coalition the users joins. As can be seen  
in Figure \ref{10UE}, though  being geographically neighboring nodes, 2 users (e.g., the 5th and 6th users) are not necessarily placed in the same coalition. This is reasonable, as indicated in Figure \ref{10_cor}, where the channel correlation between the 5th and 6th users is weak. In contrast, users 1, 9, and 10 are grouped in the same coalition due to their high channel correlation, as illustrated in Figure \ref{10_cor}.

\begin{figure}[htb]
    \centering
    \includegraphics[width=0.82\linewidth]{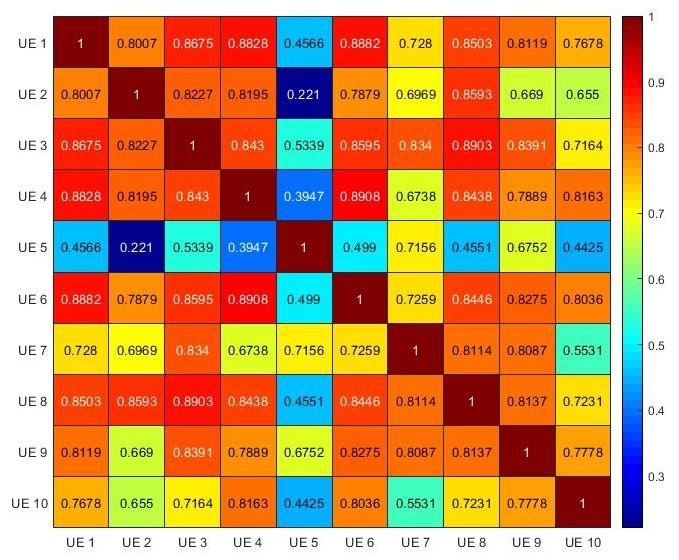}
    \caption{Channel correlation coefficient with 10 intensive users.}
    \label{10_cor_Ro}\vspace{-1.5em}
\end{figure}

A similar conclusion can be achieved for the dense distribution scenario. As shown in Figure \ref{10UE_Ro}, the actions of all users are $A_t = [3, 3, 0, 2, 0, 2, 1, 0, 1, 2]$. For instance, the channel correlation between  User 2 and User 5 is weak, and they do not choose the same coalition. In addition, User 2 prefers to join the same coalition as User 1 rather than User 10, as the channel correlation between User 2 and User 1 is stronger. This can be observed from the first row of Figure \ref{10_cor_Ro}.
The above analysis highlights the significant impact of channel correlation on coalition formation and algorithm performance, thereby demonstrating the validity of the proposed algorithm.

\vspace{-1em}\subsection{Channel estimation accuracy comparison} 
\begin{figure}[htb]\vspace{-1em}
    \centering
    \includegraphics[width=0.82\linewidth]{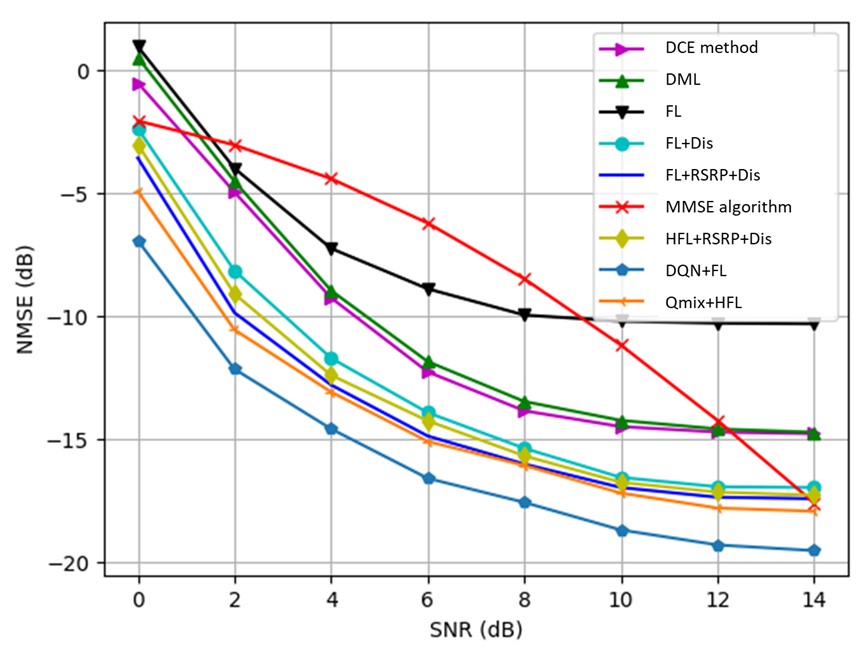}
    \caption{NMSE versus SNR under different algorithms.}
    \label{Overall}\vspace{-1em}
\end{figure}
\label{Accuracy_Coma}
The performance comparisons between the proposed scheme and several benchmark algorithms are illustrated in Figure \ref{Overall}.
% 这个图里面 除了  ``DQN+FL"   and ``Qmix+HFL"， 其他都是没有联盟形成的. 
Note that to show the performance gains brought by coalition formation,  except  ``DQN+FL"   and ``Qmix+HFL",  all other algorithms in Figure \ref{Overall} assume that all users form a single large FL group rather than being divided into multiple coalitions. 
The following conclusions can be achieved.  
1) Impact of SNR: It is shown that when the SNR increases, all algorithms achieve higher estimation accuracy. 2) NMSE gains from transfer learning: %``FL+RSRP+Dis" outperforms  ``FL+Dis",  ``DML", conventional ``FL", ``MMSE algorithm", and ``DCE methods" in terms of the  estimation accuracy. 
%Compared with the ``MMSE algorithm", the ``FL+Dis" and ``FL+RSRP+Dis" achieve superior performance at lower SNR. 
Compared with the ``DML", conventional ``FL", ``MMSE algorithm", and ``DCE method", the ``FL+Dis" and ``FL+RSRP+Dis"   improve channel estimation accuracy by $33\% \sim 50\%$. 
 Therefore, we conclude the integration of transfer learning significantly enhances NMSE performance.
This is because the transferred model has already captured a portion of the underlying channel features, and subsequent training further refines the model, enhancing estimation accuracy.  Besides, it can be seen that ``FL+RSRP+Dis"      outperforms  ``FL+Dis"  in channel estimation accuracy.  This  is because,  compared to strategies that consider only distance similarity,  ``FL+RSRP+Dis" integrates both RSRP and distance similarity enables the usage of more accurate  similar channel estimation neural network parameters  for transfer, as illustrated in \eqref{am''}  and  \eqref{r_t}. 
\begin{figure}[htb]
    \centering
    \includegraphics[width=0.98\linewidth]{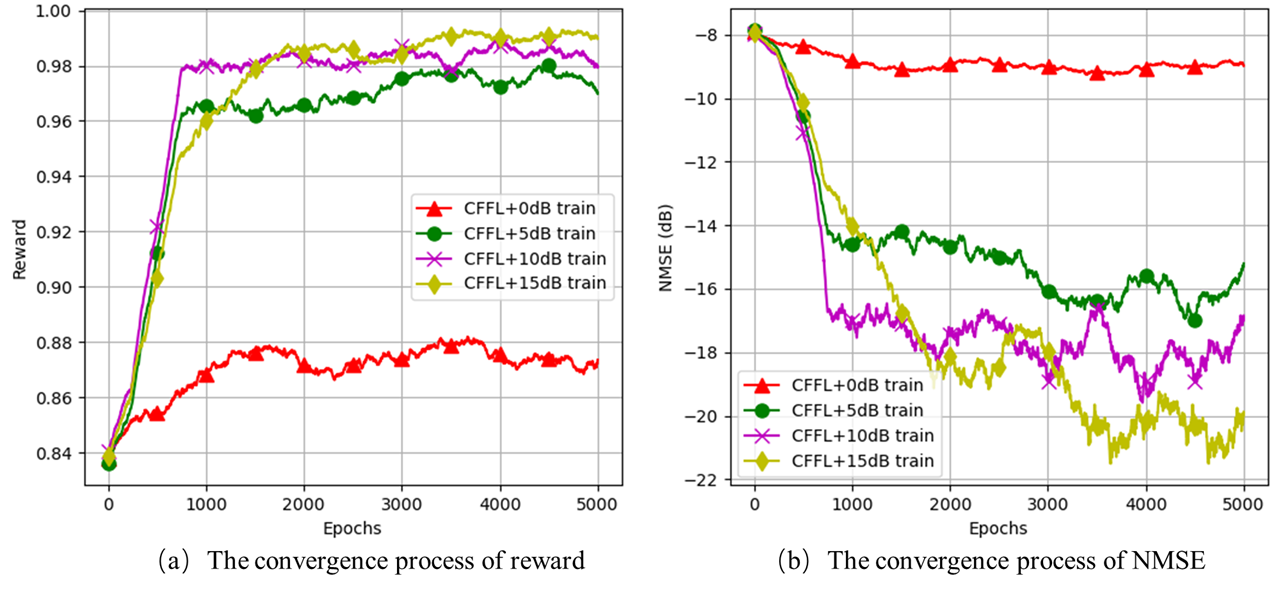}
    \caption{The performance of DQN-based algorithm at different SNR vs epochs.}
    \label{DQN_SNR}\vspace{-1em}
\end{figure}
3) HFL-based Algorithm: The HFL-based algorithm (``HFL+RSRP+Dis") outperfroms other benchmark methods by at least 5\% but slightly underperforms compared to homogeneous federated learning (``FL+RSRP+Dis") by 3\%. This is due to the slight reduction in feature extraction capability in the distilled small model compared to the original large model. However, model distillation significantly reduces computational and hardware requirements for users  (detailed numerical analyses are presented in Section \ref{FL_HFL_Coma}).  4) Impact of reinforcement learning:  The introduction of reinforcement learning algorithms improves federated learning performance. While ``Qmix+HFL", a distributed decision-making algorithm, shows   slight performance degradation compared to the  ``DQN+FL",  its  NMSE performance   ranks just below that of ``DQN+FL". Additionally, in the SNR range of 0 to 2 dB, Qmix+HFL achieves approximately a 30\% improvement in NMSE compared to ``HFL+RSRP+Dis".

\subsection{Impact of training SNR on performance}
Figure \ref{DQN_SNR}(a) depicts the convergence process of the  DQN-enabled federated learning,  where the training SNR (i.e., the SNR of the sample data) is set to 0 dB, 5 dB, 10 dB, and 15 dB, respectively. It is seen that rewards improve as SNR increases, which is reasonable. Figure \ref{DQN_SNR}(b) depicts how NMSE varies with epochs during DQN training. The performance of CFFL is much better compared to ``FL+RSRP+Dis" in Figure \ref{Overall}. For example, when SNR$=5$ dB, the NMSE achieved by the proposed algorithm is around $-16$ dB. Notably, compared  with ``FL+RSRP+Dis" and ``FL+Dis" (all users form one large   FL group, instead of being divided into different groups), our proposed CFFL brings NMSE gains of 18\% and 28\%, respectively. This is because with the reinforcement learning method, the coalition can more reasonably divide the users into different FL groups. In other words,  users experiencing similar channel fading are grouped into a coalition; thus, enhancing the effectiveness of   federated learning. Consequently, this results in a more accurate global model for each FL group and improves channel estimation accuracy.

\subsection{Performance comparison of HFL and FL, DQN and Qmix}
\begin{figure}[htb]
    \centering
    \includegraphics[width=1\linewidth]{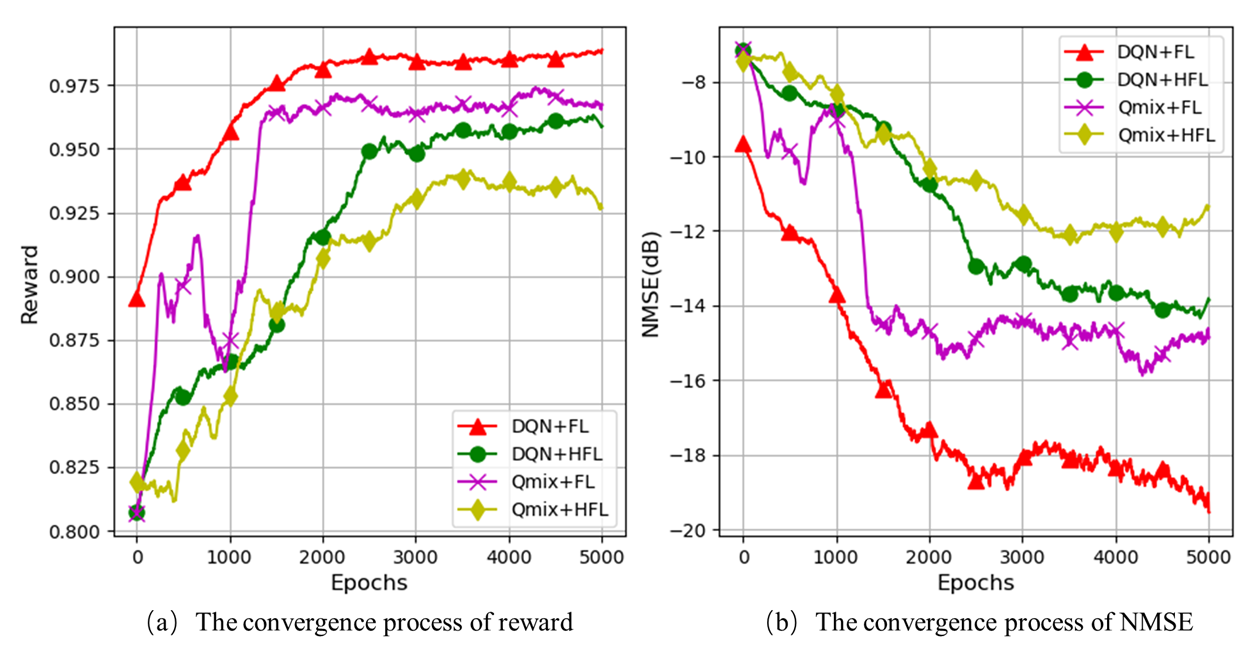}
    \caption{Rewawrd vs Epochs under different algorithms.}
    \label{DQN_Qmix_FL_HFL}
\end{figure}

Figure \ref{DQN_Qmix_FL_HFL} compares the channel estimation performance resulting from the formation of federated learning user coalitions guided by DQN and Qmix, as well as by homogeneous and heterogeneous federated learning methods. We set SNR$=5$ dB. The performance of federated learning based on Qmix, i.e., ``Qmix+FL", is about 97.5\% of that achieved with DQN, indicating a slight decrease in reward. This is because Qmix, as a distributed algorithm, only utilizes its local observation information and its own Q-network in the neural network executing phase,  leading to a performance degradation compared to the centralized DQN model. 
\begin{figure}[htb]\vspace{-1em}
    \centering
    \includegraphics[width=0.95\linewidth]{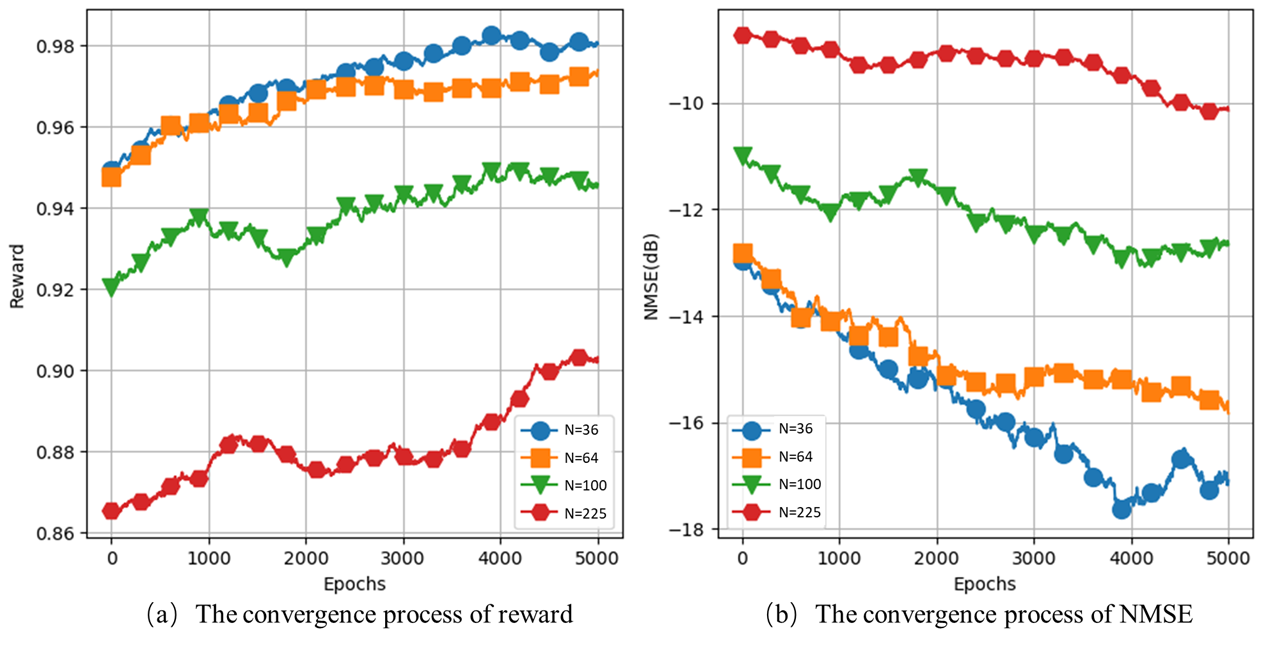}
    \caption{The performance of DQN-based algorithm at different sizes of RIS vs epochs.}
    \label{Size}\vspace{-2em}
\end{figure}
Meanwhile, the performance of HFL is about 96.8\% of that of homogeneous federated learning algorithms. However, as aforementioned, a slight decline in channel estimation performance  is compensated by significant reduction in information exchange, local user model complexity, and an improvement in user privacy security.

\vspace{-0.5em}\subsection{Impact of RIS size on performance}
Figure \ref{Size}  illustrates the performance of the DQN-based algorithm as the RIS size varies. The figure shows that the accuracy of channel estimation decreases with an increasing number of RIS elements. 
For example, at epoch 4000, the NMSE decreases by 11.4\%, 28\%, and 45.1\% for RIS  numbers of 64, 100, and 225, respectively, compared to a element number of 36. This decline occurs because changes in RIS size do not affect the signal dimensionality at the users, while the cascaded channel matrix grows larger in dimension. Given that the dimensionality of the received signal, the amount of sample data, and the network structure remain the same, inferring cascaded channels of larger dimensions (i.e., network outputs) becomes more challenging, leading to a degradation in performance.

\subsection{Impact of neural network size on the channel estimation accuracy}
\begin{figure}[htb]
    \centering
    \includegraphics[width=1\linewidth]{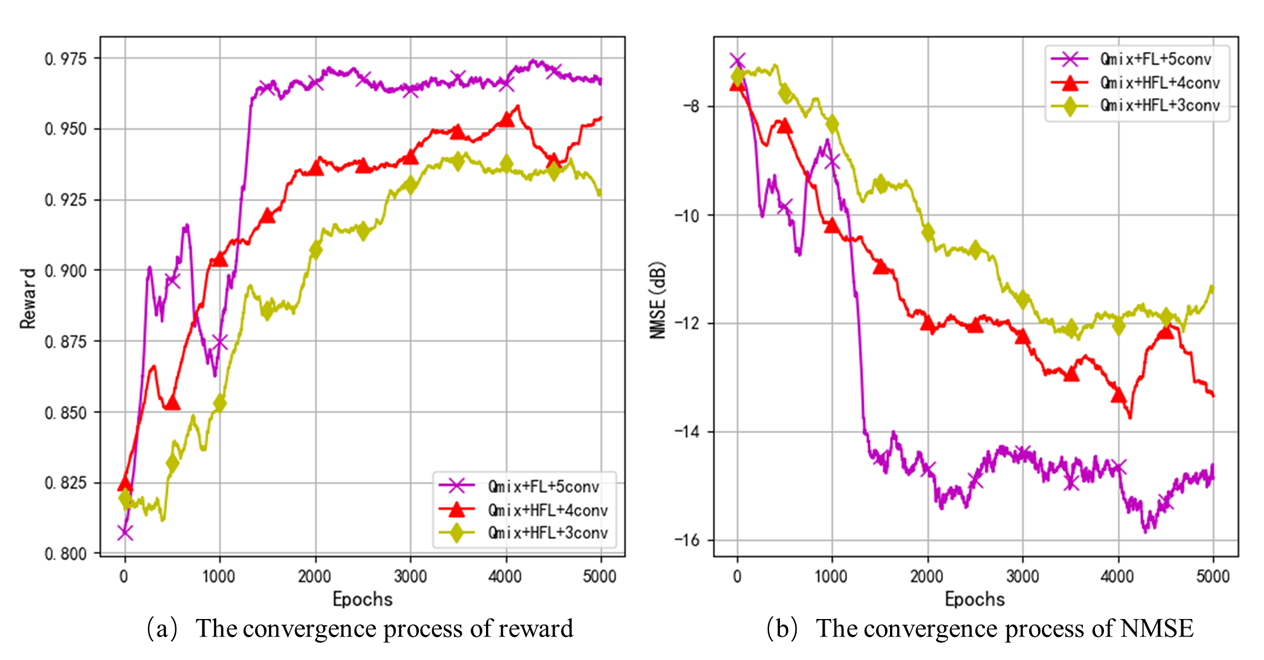}
    \caption{The performance of Qmix-based algorithm at different network layer vs epochs.}
    \label{Qmix_HFL_conv}
\end{figure}

Figure \ref{Qmix_HFL_conv} shows how the number of neural network layer affects the channel estimation accuracy.  Take the heterogeneous Qmix algorithm as an example.  The results indicate that as the number of layers in the heterogeneous network decreases, channel estimation accuracy also declines.   
For instance, at epoch 5000, the NMSE decreases by 14.1\% for convolutional layer numbers of 4 (i.e., ``Qmix+HFL+4conv") and by 25\% for a layer number of 3 (i.e., ``Qmix+HFL+3conv"), compared to a convolutional layer number of 5 (i.e., ``Qmix+FL+5conv").
It is because of the reduced inference capability due to the decrease of the neural network layer. Therefore, it is important to select an appropriate  neural network layer by balancing estimation accuracy with the hardware capabilities of the deployed devices.

\subsection{Model complexity comparison between FL and HFL}\label{FL_HFL_Coma}
Model complexity  is compared  between HFL and homogeneous federated learning.  The time taken to complete each episode is averaged over 10 episodes.  In HFL, users generate a small (distilled) local model, while the central server utilizes a larger global model. In contrast, homogeneous FL employs large neural network models for both users and the central server, resulting in significant differences in model complexity. As shown in Table \ref{Model complexity}, HFL 
 reduces computational overhead by approximately 16\%.
% \begin{figure}[htbp]
%     \centering
%     \includegraphics[width=0.95\linewidth]{figuresEH/Model_complexity1.png}
%     \caption{Model complexity comparison between HFL and homogeneous FL. The number of users is 10. The model of the GPU is RTX 1660S.}
%     \label{Model complexity}
% \end{figure}
\newcolumntype{P}{>{\centering\arraybackslash}p{4.6cm}}
\newcolumntype{Y}{>{\centering\arraybackslash}c}
\begin{table}
% \tiny
\caption{Model complexity comparison between HFL and homogeneous FL. The number of users is 10. The model of the GPU is RTX 1660S.}
\label{Model complexity}
\resizebox{1.0\linewidth}{!}{
\begin{tabular}{|Y|P|P|c|}
\hline
\textbf{Network} & \textbf{FL} & \textbf{HFL} & Improvement \\
\hline
Total parameter  & 8,428,416 (5 convolutional layers and 1 fully connected layer) & 8,409,856 (3 convolutional layers and 1 fully connected layer) & 0.22\% \\
\hline
Flops & 0.84882272G (1 big local model) & 0.695730176G (1 distilled local model) & 18.04\% \\
\hline
Total Flops & 9.337044992G (1 big global model and 10 big local models) & 7.806124032G (1 big global model and 10 small local models) & 16.40\% \\
\hline
Time for 1 Epoch (s) & 1.05349256 & 0.893589525 & 15.18\% \\
\hline

\end{tabular}
}\vspace{-2em}
\end{table}

\vspace{-0.5em}\subsection{Communication overhead comparison}
\begin{table*}[htp]
\caption{Comparison of communication interactions}\label{interactions}
\centering 
\resizebox{\textwidth}{!}{%
\begin{tabular}{|c|c|c|c|}
\hline
{} &  \textbf{One Round of FL} & \textbf{One Epoch of CFFL}& \textbf{\textbf{The total amount of interaction data required until convergence}} \\ \hline

			DQN + FL & $2K W_{FL}$ &  $E \left(2K W_{FL}\right) + K [\left( O + A\right) + P]$  &  $T_1  \left\{ E \left(2K W_{FL}\right) + K [\left( O + A\right) + P] \right\}$  \\ \hline
			
			DQN + HFL   & $2K W_{HFL} $ & $E \left(2K W_{HFL} \right) + K [\left( O + A\right) + P] $ &$T_2  \left\{ E \left(2K W_{HFL} \right) + K [\left( O + A\right) + P] \right\}$\\ \hline
			
			QMIX + FL    & $2K W_{FL}$ & $E  \left(2K W_{FL}\right) + K [\left( O + (C_{max}+1)\right) + (C_{max}+1)]$ &$T_3  \left\{ E  \left(2K W_{FL}\right) + K [\left( O + (C_{max}+1)\right) + (C_{max}+1)] \right\}$ \\ \hline
			
			QMIX + HFL &  $2K W_{HFL}$ &  $E  \left(2K W_{HFL}\right) + K [\left( O + (C_{max}+1)\right) + (C_{max}+1)]$ &$T_4  \left\{ E  \left(2K W_{HFL}\right) + K [\left( O + (C_{max}+1)\right) + (C_{max}+1)] \right\}$\\ \hline

\end{tabular}%
}\vspace{-1em}
\end{table*}

The comparison of communication interactions among the four algorithms  of CFFL  is listed in Table \ref{interactions}. Specifically, $K$ denotes the number of users, $W_{FL}$ denotes the number of model parameters to be transmitted by a single user in homogeneous FL, $W_{HFL}$ denotes the number of model parameters to be transmitted by a single user in HFL.
Consequently, the amount of    interactions for DQN-based  FL in a round is  $2K W_{FL}$. Let $E$ denote  the number of rounds to be performed in each epoch, $O$  represent  the amount of environmental information  observed by the user, $A$ indicate the amount of action information transmitted by the user, and $P$ denote the policy information provided by the executive network. Then, the total amount of communication interactions for DQN-FL used in an epoch is given by   $E \left(2K W_{FL}\right) + K [\left( O + A\right) + P]$. Let $ T_1, T_2, T_3, T_4 $ represent the number of converged epochs for ``DQN+FL", ``DQN+HFL", ``Qmix+FL",  ``Qmix+HFL", respectively. Then the total amount of communication interactions for DQN-FL is given by $T_1 \cdot \left\{ E \left(2K W_{FL}\right) + K [\left( O + A\right) + P] \right\}$. The total communication interactions for the other algorithms can be calculated in a similar manner.  It can be observed from Section III-C that the amount of communication interaction is primarily determined by the number of model parameters, specifically $W_{HFL}$ and $W_{FL}$.  
The difference in the number of parameters between FL and HFL in one round is given by $2KW_{FL} - 2KW_{HFL}$, where $K = 10, W_{FL}=8.428M, W_{HFL}=8.409M$ under the scenario settings. 
For the difference in communication interactions between DQN and Qmix, it mainly depends on the dimensions of the action size that needs to be determined and the dimension of the vector executing the policy, specifically $A= (C_{max}+1)\,K=40, P=40$, where $C_{max}=3$ is the maximum number of coalitions. This difference becomes larger as the number of users increases.  Thus, the difference in the amount of algorithmic communication interactions becomes signigcantly large after $E$ rounds and $T$ epochs of iterations.

\vspace{-0.5em}\section{Conclusion}
A coalition formation-guided HFL channel estimation method was proposed.  In particular, we utilized Qmix to implement  distributed reinforcement learning for coalition formation,  which enhancing the grouping of FL users. Additionally,  Qmix facilitates autonomy and cooperation among users in the coalition, where each user can independently make decisions based on local information, while considering the interests of the entire coalition. This approach significantly reduces the overhead associated with information exchange compared to the centralized DQN method. Furthermore, the HFL-based approach was proposed to reduce the size of local neural network models for individual users.  Numerical results showed that although HFL   slightly reduces channel estimation accuracy, it significantly lowers computational cost and complexity, and reduces training duration by approximately  15\%. More importantly, HFL transmits only a subset of model parameters, reducing the amount of data transmitted and better protecting local user privacy compared to homogeneous FL.
Simulations revealed that, proposed method   reduces the end-user computational overhead    by 16\%, improves data privacy, and enhances    channel estimation accuracy by 20\%. Of note in our context, the HFL (without coalition formation) and the DQN-enabled homogeneous FL serve as the  lower and  upper performance bounds, respectively. The Qmix-enabled algorithm outperforms the lower performance bound baseline by 25\% while maintaining an accuracy loss of less than 4\% compared to the upper-performance bound. 
 
\renewcommand{\thesection}{\Alph{section}} % 如果需要，重定义章节编号样式为大写字母
\begin{appendices}
    \section{}\vspace{-0.2em} % 第一个附录

Considering the altruistic cooperation among  UEs, the utility function of UE $j$ is defined as
\begin{equation}
U=u\left(a_{m, j}\right)+\sum_{i \in S_m \setminus \{j\}} u\left(a_{m, i}\right)+\sum_{i \in S_n} u\left(a_{n, i}\right),
\end{equation}
where $S_m$ represents the coalition that   UE $j$ leaves,   $S_n$ represents the new coalition that   UE $j$ chooses to join, $S_m \setminus \{j\}$ represents the coalition $S_m$ after removing  UE $j$. The potential  function is defined as  
%The potential energy function is defined as follows
\begin{equation}
\varphi=\sum_{j \in S_k, S_k \in S} u\left(a_{k, j}\right).
\end{equation}

If the UE unilaterally changes its coalition choice from $S_m$ to $S_n$, then the UE utility function changes by
\begin{equation}
\begin{scriptsize}
\begin{aligned}
U^{\prime}-U & =\left\{u^{\prime}\left(a_{n, j}\right)+\sum_{i \in S_n^{\prime} \setminus \{j\}} u^{\prime}\left(a_{n, i}\right)+\sum_{i \in S_m^{\prime}} u^{\prime}\left(a_{m, i}\right)\right\} \\
&-\left\{u\left(a_{m, j}\right)+\sum_{i \in S_m \setminus \{j\}} u\left(a_{m, i}\right)+\sum_{i \in S_n} u\left(a_{n, i}\right)\right\} \\
& =\underbrace{\left\{u^{\prime}\left(a_{n, j}\right)-u\left(a_{m, j}\right)\right\}}_{\footnotesize{\textcircled{\scriptsize{1}}}\footnotesize}+\underbrace{\left\{\sum_{i \in S_n^{\prime} \setminus \{j\}} u^{\prime}\left(a_{n, i}\right)-\sum_{i \in S_n} u\left(a_{n, i}\right)\right\}}_{\footnotesize{\textcircled{\scriptsize{2}}}\footnotesize}\\
&+\underbrace{\left\{\sum_{i \in S_m^{\prime}} u^{\prime}\left(a_{m, i}\right)-\sum_{i \in S_m \setminus \{j\}} u\left(a_{m, i}\right)\right\}}_{\footnotesize{\textcircled{\scriptsize{3}}}\footnotesize} 
\end{aligned}
\end{scriptsize}
\end{equation}
where $S_m^{\prime}$ represents the coalition that UE $j$ leaves, $S_n^{\prime}$ represents the new coalition that UE $j$ chooses to join.

On the other hand, since the UE unilaterally changes its coalition selection, the change in the potential function can be calculated as 
% the change of the terrain energy function is as
\begin{equation}
\begin{tiny}
\begin{aligned}
\varphi^{\prime}-\varphi= & \left\{\sum_{j \in S_k, S_k \in S^{\prime}} u^{\prime}\left(a_{k, j}\right)\right\}-\left\{\sum_{j \in S_k, S_k \in S} u\left(a_{k, j}\right)\right\} \\
= & \underbrace{\left\{u^{\prime}\left(a_{n, j}\right)-u\left(a_{m, j}\right)\right\}}_{\footnotesize{\textcircled{\scriptsize{4}}}\footnotesize} +\underbrace{\left\{\sum_{i \in S_n^{\prime} \backslash\{j\}} u^{\prime}\left(a_{n, i}\right)-\sum_{i \in S_n} u\left(a_{n, i}\right)\right\}}_{\footnotesize{\textcircled{\scriptsize{5}}}\footnotesize}\\
&+\underbrace{\left\{\sum_{i \in S_m^{\prime}} u^{\prime}\left(a_{m, i}\right)-\sum_{i \in S_m \backslash\{j\}} u\left(a_{m, i}\right)\right\}}_{\footnotesize{\textcircled{\scriptsize{6}}}\footnotesize} \\
& +\underbrace{\left\{\sum_{i \in S_{k,}, S_k \in S^{\prime} \backslash\left\{S_n^{\prime}, S_m^{\prime}\right\}} u^{\prime}\left(a_{k, i}\right)-\sum_{i \in S_k, S_k \in S \backslash\left\{S_n, S_m\right\}} u\left(a_{k, i}\right)\right\}}_{\footnotesize{\textcircled{\scriptsize{7}}}\footnotesize}
\end{aligned}
\end{tiny} \label{proof}
\end{equation}

Since UE $j$ unilaterally changes its coalition choice, it will only affect the coalition from which UE $j$  is departing and to which it is joining, namely $S_m,S_n,S_m^{\prime},S_n^{\prime}$. Therefore, \normalsize{\textcircled{\scriptsize{7}}}\normalsize\enspace in (\ref{proof}) is 0.
In addition, ${\textcircled{\scriptsize{1}}}={\textcircled{\scriptsize{4}}}$,  ${\textcircled{\scriptsize{2}}}={\textcircled{\scriptsize{5}}}$,  ${\textcircled{\scriptsize{3}}}={\textcircled{\scriptsize{6}}}$. Hence, we have 
% the following equation holds
% \begin{equation}
% \begin{footnotesize}
% \begin{aligned}
% \sum_{i \in S_{k,}, S_k \in S^{\prime} \backslash\left\{S_n^{\prime}, S_m^{\prime}\right\}} u^{\prime}\left(a_{k, i}\right)-\sum_{i \in S_k, S_k \in S \backslash\left\{S_n, S_m\right\}} u\left(a_{k, i}\right) = 0
% \end{aligned}
% \end{footnotesize}
% \end{equation}
\vspace{-0.5em}\begin{equation}
\varphi^{\prime}-\varphi = U^{\prime}-U. \label{eqp}\vspace{-0.3em}
\end{equation}
\eqref{eqp} indicates  that when UE $j$ ($\forall j$) independently changes its coalition selection, the variation in the utility function is equal to that of the potential function. This implies that  the coalition formation game is   an exact potential game, which indicates that there is at least one pure strategy Nash equilibrium \cite{potentialgame}. Consequently, no UE can improve its utility by unilaterally changing its coalition.    Since the potential function and the objective function are positively and linearly related,  maximizing the potential function is equivalent to maximizing the utility of the coalition formation of the whole network. This concludes the proof of \textbf{Theorem 1}. 

% \section{} % 第二个附录
% In simulations, the center frequency of the subcarriers used for OFDM is set to 2.6GHz, a common frequency band for 5G applications. The user terminal's mobility speed is assumed to be approximately 60 km/h (i.e., 16.67 m/s). The signal bandwidth $B$ is set to 100 MHz, and the number of subcarriers $N_s$ is  fixed to 1024, resulting in a subcarrier spacing $\Delta f = B / N_s$.  The channel coherence time is given by $T_c = 1 / f_d=c / (v \times f_0)=6.92 $ ms, where $f_d$ represents the maximum Doppler frequency. An OFDM symbol duration can be expressed as $T_s=1 / \Delta f=100 \mu$s. The length of the pilot signal used is 128, then the time available to record the channel information is $T=128 \times T_s = 1.28 $ ms which is much smaller than  $  T_c=6.92 $ ms.  Therefore, the proposed algorithm can be applied to channel estimation in most scenarios.

\end{appendices}

\bibliography{references}
\bibliographystyle{IEEEtran}
\balance

% \newpage
 
% \begin{IEEEbiography}
% [{\includegraphics[width=1in,height=1.25in, clip,keepaspectratio]{figuresEH/Nan Qi.jpg}}]{Nan Qi} (Senior Member, IEEE) received the B.Sc. and Ph.D. degrees in communications engineering from Northwestern Polytechnical University, China, in 2011 and 2017, respectively. She is
% also a Postdoctoral Scholar with the KTH Royal Institute of Technology, Sweden. She is currently an Associate Professor with the Department of Electronic Engineering, Nanjing University of
% Aeronautics and Astronautics, China. Her research interests include optimization of wireless communications, opportunistic spectrum access, learning theory, and game theory.
% \end{IEEEbiography}

% \begin{IEEEbiography}[{\includegraphics[width=1in,height=1.25in, clip,keepaspectratio]{figuresEH/Haoxuan Liu.jpg}}]{Haoxuan Liu} received the B.S. degree in communications and information systems from the College of Electronic and Information Engineering, Nanjing University of Aeronautics and Astronautics (NUAA), Nanjing, China, in 2022, where he is currently pursuing the M.A. degree. His research interests include intelligent reflecting surface, game theory, and convex optimization.
% \end{IEEEbiography}


% \vfill

\end{document}